\documentclass[12pt]{article}

\usepackage{scicite}
\usepackage{graphicx}
\usepackage{amssymb}
\usepackage{xcolor}
\usepackage{hyperref}
\usepackage[normalem]{ulem}
\usepackage{times}
\usepackage{textcomp}

\newenvironment{sciabstract}{%
\begin{quote} \bf}
{\end{quote}}

\newcounter{lastnote}

\date{}
\newcommand{\ccm}{\rm ~cm^{-3}}
\newcommand{\kms}{\rm ~km~s^{-1}}
\newcommand{\msun}{\rm ~M_\odot}
\newcommand{\ergs}{\rm ~erg~s^{-1}}
\newcommand{\wl}{\lambda~}
\newcommand{\wll}{\lambda\lambda~}

\begin{document} 

\begin{center}
\textbf{\Large{Emission lines due to ionizing radiation from a compact object in the remnant of Supernova 1987A} }
\\
\vspace{1cm}
C.\ Fransson,$^{1\ast}$ M.\ J.\ Barlow,$^{2\dag}$ P.\ J.\ Kavanagh,$^{3,4\dag}$ J.\ Larsson,$^{5\dag}$
O.\ C.\ Jones,$^{6}$ \\
B.\ Sargent, $^{7,8}$   M.\ Meixner,$^{9}$ P.\ Bouchet,$^{10}$
T.\ Temim,$^{11}$
G.\ S.\ Wright,$^{6}$ \\
J.\ A.\ D.\ L.\ Blommaert,$^{12}$ N.\ Habel,$^{9}$ A.\ S.\ Hirschauer,$^{7}$ J.\ Hjorth, $^{13}$  
L.\ Lenki\'{c},$^{14}$   \\
T.\ Tikkanen,$^{15}$ R.\ Wesson,$^{16}$ A.\ Coulais,$^{17,10}$ O.\ D.\ Fox,$^{7}$ R.\ Gastaud,$^{18}$ \\ A.\ Glasse,$^{6}$  J.\ Jaspers,$^{3,4}$ O.\ Krause,$^{19}$ R.\ M.\ Lau,$^{20}$ O.\ Nayak,$^{21}$ \\ A.\ Rest,$^{7,8}$ 
L.\ Colina,$^{22}$  E.\ F.\ van Dishoeck,$^{23,24}$ M.\ G\"udel$^{25,19,26}$ Th.\ Henning,$^{19}$ \\ P.-O.\ Lagage,$^{10}$  G.\ \"Ostlin,$^{1}$  T.\ P.\ Ray,$^{4}$ B.\ Vandenbussche$^{27}$  \\
\vspace{0.5cm}

\scriptsize{
$^{1}$Department of Astronomy, Stockholm University, The Oskar Klein Centre,  
AlbaNova, SE-106 91 Stockholm, Sweden \\
$^{2}$Department of Physics and Astronomy, University College London, 
London WC1E 6BT, UK \\
$^{3}$Department of Experimental Physics, Maynooth University,
 Maynooth, Co Kildare, Ireland \\
$^{4}$Astronomy \& Astrophyics Section,
School of Cosmic Physics, Dublin Institute for Advanced Studies, Dublin 2, Ireland \\
$^{5}$Department of Physics, KTH Royal Institute of Technology, The Oskar Klein Centre,
AlbaNova, SE-106 91 Stockholm, Sweden \\
$^{6}$
UK Astronomy Technology Centre, Royal Observatory, 
Blackford Hill, Edinburgh, EH9 3HJ, UK \\
$^{7}$
Space Telescope Science Institute,
Baltimore, MD 21218, USA \\
$^{8}$Department of Physics and Astronomy, The Johns Hopkins University,
Baltimore, MD 21218, USA \\
$^{9}$
Jet Propulsion Laboratory, California Institute of Technology, Pasadena, CA 91109, USA \\ 
$^{10}$Universit\'e Paris-Saclay, Université Paris Cité, Commissariat à l'énergie atomique et aux énergies alternatives (CEA),
Centre National de la Recherche Scientifique (CNRS), Astrophysique, Instrumentation, Modélisation (AIM),
L'Orme des Merisiers
91191 St Aubin, France \\
$^{11}$Department of Astrophysical Sciences, Princeton University, Princeton, NJ 08544, USA \\
$^{12}$Astronomy and Astrophysics Research Group, Department of Physics and Astrophysics, \\ Vrije Universiteit Brussel, 
B-1050 Brussels, Belgium \\
$^{13} $Dark Cosmology Centre, Niels Bohr Institute, 
University of Copenhagen,
2200 Copenhagen, Denmark \\
$^{14}$Stratospheric Observatory for Infrared Astronomy  Science Center, Universities Space Research Association, \\ NASA Ames Research Center, Moffett Field, CA 94035, USA \\
$^{15}$School of Physics \& Astronomy, Space Research Centre, Space Park Leicester, University of Leicester, Leicester LE4~5SP, UK \\
$^{16}$School of Physics and Astronomy, Queen’s Buildings, Cardiff University, Cardiff, CF24 3AA, UK \\
$^{17}$Laboratoire d'Etudes du Rayonnement et de la Matière en Astrophysique et Atmosphères (LERMA), 
Observatoire de
Paris, Paris Sciences \& Lettres (PSL) Research University,
French National Centre for Scientific Research (CNRS), Sorbonne Universit\'e, Paris, France \\
$^{18}$
Universit\'e Paris-Saclay, Commissariat à l'énergie atomique et aux énergies alternatives (CEA), 
\\ Detectors, Electronics and Computing for Physics (DEDIP), 91191, Gif-sur-Yvette, France \\
$^{19}$Max-Planck-Institut für Astronomie,
K\"onigstuhl 17, D-69117 Heidelberg, Germany \\
$^{20}$National Science Foundation's National Optical-Infrared Astronomy Research
Laboratory, Tucson, AZ 85719, USA \\
$^{21}$NASA Goddard Space Flight Center, Greenbelt, MD 20770, USA \\
$^{22}$Centro de Astrobiolog\'{\i}a, Consejo Superior de Investigaciones Científicas -  Instituto Nacional de Técnica Aeroespacial
(CSIC-INTA), Torrej\'on de Ardoz,
E-28850, Madrid, Spain \\
$^{23}$Max-Planck Institut f\"ur Extraterrestrische Physik,
D-85748, Garching, Germany \\
$^{24}$Leiden Observatory, 2300 RA Leiden, The Netherlands \\
$^{25}$Dept. of Astrophysics, University of Vienna, A-1180 Vienna, Austria \\
$^{26}$Institute for Particle Physics and Astrophysics, Eidgenössische Technische Hochschule (ETH) Z\"urich, 8093 Z\"urich, Switzerland \\
$^{27}$Institute of Astronomy, Katholieke Universiteit Leuven, 3001 Leuven, Belgium
\\
$^\ast$To whom correspondence should be addressed; E-mail:  claes@astro.su.se.
\\
$^\dag$These authors contributed equally to this work.
}
\end{center}
\newpage

\baselineskip24pt

\newpage
\begin{sciabstract}
The nearby Supernova 1987A was accompanied by a burst of neutrino emission, which indicates that a compact object (a neutron star or black hole) was
formed in the explosion. There has been no direct observation of this compact
object. In this work, we observe the supernova remnant with JWST spectroscopy finding
narrow infrared emission lines of argon and sulphur. The line emission is spatially unresolved and blueshifted in velocity relative to the supernova rest frame. We interpret the lines as gas illuminated by a source of ionizing
photons located close to the center of the expanding ejecta.  Photoionization models show that the line ratios are consistent with ionization by a cooling neutron star or pulsar wind nebula. The velocity shift could be evidence for a neutron star natal kick.

\end{sciabstract}

Supernova 1987A (SN 1987A)  in the Large Magellanic Cloud is the most recent supernova (SN) in the Local Group of galaxies, so it has been extensively studied. It originated from the collapse of a blue supergiant
star of 15 to 20 solar masses ($\msun$). Observations taken 35 years after the explosion show that the remnant of SN 1987A contains several distinct components (Fig.~\ref{fig:components})
\cite{McCray2016}. The inner ejecta containing the synthesized heavy elements produced in the explosion, are now interacting with an equatorial ring (ER) and the surrounding circumstellar medium  \cite{Larsson2023}. The ER is thought to have originated  $\sim 20,000$ years before the explosion, probably from  a binary merger experienced by the progenitor star \cite{Morris2007,Morris2009}. The inner ejecta produce 
broad emission lines with velocities of 2000 to 4000 $\kms$, powered by radioactive decay of ${}^{44}$Ti nuclei  and x-rays generated by the interaction with the ER \cite{McCray2016}. 

A compact object, such as a neutron star (NS) or black hole (BH), is expected to be
located in the central region, but has not been observed. 
A neutrino burst was detected coinciding with SN 1987A \cite{Alekseev1987,Bionta1987,Hirata1987}; its duration indicates that the compact object is a NS \cite{Burrows1988}.
Ionizing photons emitted by such a NS have been predicted to excite emission lines from heavy elements in the ejecta with velocities 100 to 1000 $\kms$ \cite{CF1992}. In this study we searched for those emission lines.

\section*{Infrared spectroscopy of SN1987A}
We observed the SN 1987A remnant on 16 July 2022 (day 12,927 after the SN explosion) 
using 
the JWST with the Mid-Infrared Instrument/Medium Resolution Spectrometer \\ (MIRI/MRS) and the Near-Infrared Spectrograph/Integral Field Unit (NIRSpec/IFU). Both instruments produce data cubes, which contain a spectrum for each spatial pixel \cite{MM}. 

The data cubes exhibited an excess of emission in the central pixels of the ejecta (Fig. 2)
at observed peak wavelengths of $4.5293 \pm 0.0003$~{\textmu}m in the NIRSpec data and at $6.98606 \pm 0.00003$~{\textmu}m in the MRS data,
with the latter line being much brighter. 
We identify these lines as due to [Ar~{\sc vi}] and [Ar~{\sc ii}], which have rest wavelengths 4.52922 and $6.985274$~{\textmu}m, respectively. Figure 2, A to P shows velocity slices of the MRS data cube centered on the wavelength of the [Ar~{\sc ii}] line and Fig. 2, Q to R shows slices of the NIRSpec cube, containing the [Ar~{\sc vi}] line.
The [Ar~{\sc ii}] line in SN 1987A has been observed previously
\cite{Arendt2016}, but the spatial and spectral resolution were insufficient to separate the emission from the ejecta  and the ER.
The JWST observations show it arises from a central source, separated from the ER.  

Fig.~\ref{fig:line_prof}A shows the [Ar~{\sc ii}] line profile from the central source, which has a narrow strong peak and a weaker broad extension to the red.   
We fitted this asymmetric line profile with 
two Gaussian functions. 
Fig. \ref{fig:line_prof_ArII_ArIV_ER} compares the line from the center of the ejecta with the corresponding [Ar~{\sc ii}] line from the ER.
The narrow emission line component in the center is displaced by $-259.6  
\kms$ with respect to the rest frame of SN 1987A, which has a heliocentric velocity +286.5 km~s$^{-1}$ \cite{Groningsson2008}, whereas
in the spectrum of the entire ER the [Ar~{\sc ii}] line shows a much smaller blueshift of $-12.2 \pm 1.6  \kms$ \cite{MM}. 
The velocity difference and spatial distribution indicate that the emission line from the central source is not scattered light from the ER.

The spectral resolution at the [Ar~{\sc ii}]  wavelength  corresponds to a full-width at half-maximum (FWHM) of $87.7 \pm 1.1 \kms$ \cite{Jones2023}. We measure that the narrower ejecta line component has a FWHM $= 121.9 \pm 1.3 \kms$, so it is marginally spectrally resolved. The emission from the entire ER has a FWHM$ = 286 \kms$, and the broader 
component has a value of $ = 362.4 \kms$. 
Table \ref{table:1} lists the velocity offsets, FWHMs and luminosities that we measured for the emission lines from the central ejecta.
\begin{table}[!ht]
\centering
\caption{\textbf{Parameters for emission line components from the central ejecta.} Luminosities calculations adopt a distance to SN~1987A of 49.6~kiloparsecs \cite{Pietrzynski2019}. 
No broad component was required for the $[$Ar~{\sc iii}$]$, $[$S\phantom{r}~{\sc iv}$]$, and $[$S\phantom{r}~{\sc iii}$]$ lines (fig. \ref{fig:line_prof_SIII_SIV_FeII_SM}). The $[$Ar~{\sc vi}$]$ lines were measured in the NIRSpec data; all other lines are from the MIRI data. Uncertainties are $1\sigma$ 
errors on the fitting parameters; upper limits are 3$\sigma$.  Dashes indicate not applicable.
}
\vspace{2mm}
\footnotesize{
\addtolength{\tabcolsep}{-0.2em}
\begin{tabular}{|l l| l l l| l l l|} 
 \hline
& \multicolumn{1}{c}{ }&\multicolumn{3}{c}{Narrrow component }&\multicolumn{3}{c|}{Broad component } \\  
 Line &Wavelength & Velocity offset & FWHM  &Luminosity& Velocity offset & FWHM &Luminosity \\ 
 & {\textmu}m&$\kms$&$\kms$&$10^{30} \ergs$  &$\kms$&$\kms$& $10^{30} \ergs$   \\
 \hline\hline
 &&&&& & & \\ 
\vspace{-3mm}
$[$Ar~{\sc vi}$]$ &\phantom{1}4.529&-269$\pm$24&	$<$377\phantom{.1}&3.04$\pm$0.42 &15$\pm$ 33 &$<$461\phantom{.1}&2.88$\pm$0.40\\
 &&&&& & & \\ 
\vspace{-3mm}
$[$Ar~{\sc ii}$]$ &\phantom{1}6.985&-259.6$\pm$0.4&121.9$\pm$1.3&142.9$\pm$2.7&-200.6$\pm$3.7&362.4$\pm$8.0&104.1$\pm$3.0\\
 &&&&& & & \\ 
\vspace{-3mm}
$[$Ar~{\sc iii}$]$ &\phantom{1}8.991&&&$< 3.5$&$-$&$-$&$-$\\
 &&&&& & & \\ 
\vspace{-3mm}
$[$S\phantom{r}~{\sc iv}$]$ &10.510&-234$\pm$12\phantom{.9}&145$\pm$35&2.13$\pm$0.45&$-$&$-$&$-$ \\
 &&&&& & & \\ 
\vspace{-3mm}
$[$S\phantom{r}~{\sc iii}$]$\phantom{.1}&18.713&-288$\pm$13&158$\pm$39&4.50$\pm$0.82&$-$&$-$&$-$  \\ 
 &&&&& & & \\ 

\hline
\end{tabular}
}

\label{table:1}
\end{table}

The NIRSpec spectral resolution at the [Ar~{\sc vi}] $4.529$-{\textmu}m  line is $\sim 261 \kms$. 
Fitting a similar two-component model as for the [Ar~{\sc ii}] line, we found that the peak of the [Ar~{\sc vi}] line has consistent velocity to that of [Ar~{\sc ii}], but the broad components differ from the narrow blueshifted component (Table \ref{table:1}).  
Both the [Ar~{\sc ii}] and [Ar~{\sc vi}] peaks are consistent with being spatially unresolved [the spatial resolution FWHM is $0.''350$ and $0.''212$ for [Ar~{\sc ii}] and [Ar~{\sc vi}], respectively \cite{MM}].
The position of maximum emission in  the [Ar~{\sc ii}] and [Ar~{\sc vi}] lines spatially coincide 
(within the uncertainties) with each other and 
with the previously determined center of the ER (Fig.~\ref{fig:components}), which is assumed to be the explosion site \cite{Alp2018}. 

In the MIRI data we also detected lines of [S~{\sc iv}] $10.51$~{\textmu}m and [S~{\sc iii}] $18.71$~{\textmu}m (fig. \ref{fig:line_prof_SIII_SIV_FeII_SM}). Both sulphur lines
show weak components from the central ejecta, displaced from the SN 1987A velocity by a similar amount to the argon lines. We can only place an upper limit on the luminosity of an [Ar~{\sc iii}] line at $8.991$~{\textmu}m line  (Table \ref{table:1}).
In the NIRSpec data,  we  detected the 
 [Ca~{\sc iv}] line at $3.207$~{\textmu}m  (fig. \ref{fig:line_prof_CaIV_VI_SM}) ; however, its velocity is shifted 
to the red and its maximum emission is located  slightly north of the [Ar~{\sc vi}] peak, so it might be unrelated  to the argon lines \cite{MM}.

\section*{Interpretation of the central emission}

The only narrow emission components from the ejecta that we identify are from Ar and S, which are both produced by the nuclear burning (fusion) of O and Si. This indicates that the emission arises from the inner core of the ejecta (fig. \ref{fig:abundances} ).  
Explosion models of SN 1987A \cite{WoosleyHeger2007} show that the innermost $\sim 0.07 \msun$ of ejecta is dominated by He and ${}^{56}$Ni  (the latter decays into ${}^{56}$Fe). The next zone outward, with a mass of $\sim 0.3 \msun$, contains material produced by oxygen burning -- Si, S, Ar, Ca, and, in its outer region, also O \cite{MM}. 

The same elements have been observed in young oxygen-rich supernova remnants (SNRs) from massive star explosions, including Cas A, SNR G54.1+0.3 and SNR 0540-69.3 \cite{Chevalier1979,Smith2009,Temim2010,Williams2008} (supplemenatry text). Both SNR G54.1+0.3 and SNR 0540-69.3 contain pulsar wind nebulae (PWNe), powered by the spin-down power from a rapidly rotating NS with a strong magnetic field. 
By contrast, Cas A contains a central, cooling NS  (CNS), with a surface temperature $\sim 1.8 \times 10^6$ K \cite{Ho2009} without a PWN. The emission lines in Cas A are mainly excited by a shock in the outer parts of the SN ejecta. 

The narrow [Ar~{\sc ii}], [Ar~{\sc vi}], [S~{\sc iii}] and [S~{\sc iv}] lines that we observe from the center of SN~1987A must be excited by a source of ionizing photons or a shock wave. Potential sources include photons from a PWN generated by a NS, photons directly from a CNS, accretion onto a compact object, or shock excitation by a 
PWN shock. We also considered five other possibilities (supplementary text): ionization by 
radioactive ${}^{44}$Ti; excitation produced by  x-rays from the ejecta and circumstellar medium interaction;  a surviving companion star; reflection of the narrow line emission from the ER collision by dust; or  emission from an ingoing, reverse shock. All five possibilities were excluded.

The remaining potential explanations all include the presence of a young NS or a BH in the center. 
We consider a BH unlikely because the progenitor star of SN 1987A is thought to have had a total mass of $\lesssim 20 \msun$ and an Fe core mass of $\lesssim 2 \msun$ \cite{Sukhbold2018}. The explosion ejected at least $0.07 \msun$ of ${}^{56}$Ni \cite{Bouchet1991}. That leaves less core mass remaining than is required to form a BH, which is $\gtrsim 2.2 \msun$ \cite{Musolino2023}.
Accretion of ejecta bound to the NS could be a possibility. The estimated luminosity from this, $\sim 2 \times 10^{32} \ergs$ at 35 years \cite{Chevalier_NS_acc_1995}, is too low compared with the observed luminosity (Table \ref{table:1}), but the relevant physical parameters are  uncertain.

A hot CNS would have a temperature of $\gtrsim 10^6$ K and luminosity $\gtrsim 10^{34} \ergs$ \cite{Beznogov2021}, which is sufficient to ionize part of the inner ejecta \cite{MM}. 
An analogous case is the NS in Cas A, although it is  older and fainter than what is expected for SN 1987A. 

The energy loss from a pulsar is in the form of electrons, positrons and magnetic fields, which produce an expanding high-pressure PWN. 
High-frequency synchrotron radiation from the relativistic particles can  
ionize the ejecta \cite{CF1992}.
The luminosities and spectra of the PWNs discussed above are not consistent with the bolometric and monochromatic luminosity in 
SN 1987A,
whereas a slowly rotating and/or weakly magnetized NS is compatible with the observations \cite{Alp2018}. 
We expect an expanding PWN bubble to sweep up a dense shell of ejecta, separated from the PWN by a shock, which could also excite the lines we observed. The velocity of the dense shell would depend on the energy input from the pulsar along with the density and expansion velocity of the ejecta \cite{CF1992}. These parameters are all uncertain, but estimates for SN 1987A
produce a shell velocity of 150 to 300 $\kms$ and a PWN shock velocity of  30 to 50 $\kms$ \cite{MM}. 

\section*{Photoionization models}
To investigate these scenarios in more detail, we have updated a modelling code \cite{CF1992} and applied it to model the photoionization from a PWN and from a CNS \cite{MM}. 
The resulting emission line spectrum is primarily determined by the ionization parameter, $\xi=L_{\rm ion}/n_{\rm ion} r^2$, where $L_{\rm ion}$ is the ionizing luminosity, $n_{\rm ion}$ is the number density of ions and
$r$ is the distance to the ionizing source; it is affected to a lesser extent by the spectral form, $L(\nu)$, where $\nu$ is the frequency \cite{MM}. For the ionizing spectrum of a PWN we assume a power law, $L(\nu) \propto \nu^{-\alpha}$, with $\alpha=1.1$. This matches the ultraviolet--to--x-ray spectrum of the Crab Nebula  \cite{Lyutikov2019}, and is within the range of $0.5 < \alpha < 1.1$  previously found for x-rays in young PWNs\cite{Li_PWN_spectra_2008,Gotthelf2003}. We also investigated using steeper power laws and found that it has little effect (supplementary text). Previous work has constrained $L_{\rm ion} \lesssim 10^{36} \ergs$ at energies above 13.6 eV \cite{Alp2018}. 

For the young CNS scenario, previous models have found  
an upper limit to the thermal luminosity of $\lesssim 4 \times 10^{35} \ergs$ and a surface temperature of $\sim(1.5$ to $3)\times 10^6$ K at 35 years after the explosion \cite{Beznogov2021,Beznogov2023,Page2020}.   
Central compact objects, like the NS in Cas A, 
are thought to have carbon-dominated atmospheres \cite{Ho2009}, although alternative models producing a hard spectrum have been discussed \cite{Alford2023}.  
We adopted a carbon-dominated CNS spectrum \cite{Ho2009} with 
$L_{\rm ion} = 3 \times 10^{35} \ergs$.

The models depend on the density in the ejecta core, which is uncertain. After explosion, the mass inside the oxygen core of a 19 $\msun$ progenitor is $\sim 3.0 \msun$, excluding the material that forms the NS \cite{WoosleyHeger2007}. 
The line profiles 
indicate a core velocity of $\sim 2200 \kms$ \cite{MM}. For a uniform core density $n_{\rm ion} \approx 2.6 \times 10^{3} (A/22)^{-1} \ \ccm$ after 35 years , where $A$ is the mean atomic weight in the O-Si-S-Ar core. 
However, both observations \cite{Spyromilio1991,Li1992,Larsson2013,Matsuura2017} and  
simulations \cite{Gabler2021}, indicate that clumping and filaments 
produce regions with higher densities in this zone, whereas the Fe-rich regions can have lower density as a result of early heating by decay of ${}^{56}$Ni  \cite{MM}.
If a PWN is present it produces a dense shell (as discussed above), which is likely to be unstable and clumpy \cite{CF1992,Blondin2017}.  
We therefore assume a density of
$n_{\rm ion}=2.6 \times 10^4$ cm$^{-3}$, which is 10 times the average estimated above, in both models.

Figure ~\ref{fig:photcalc} shows the lines predicted by the models of the PWN and CNS scenarios for an ejecta composition (fig. \ref{fig:abundances}) typical of explosive oxygen burning in a SN with a $19-\msun$ progenitor star \cite{WoosleyHeger2007,MM}. 
The blueshifted  line velocities and dust effects indicate that only a fraction of the total ionizing luminosity is converted to  the observed emission lines (supplementary text). We therefore normalize the [Ar~{\sc ii}] luminosity by a covering factor, chosen to match the observed luminosity, and scale the other lines to their relative model luminosities \cite{MM}. For consistency, we include only the luminosities of the blueshifted narrow component.

The relative luminosities of the [Ar~{\sc ii}] $6.985$-{\textmu}m and the [Ar~{\sc vi}] $4.529$-{\textmu}m lines are reproduced in both models (Fig. \ref{fig:photcalc}), with this assumed density and $\xi \approx 0.3$ \cite{MM}.
Both models predict other high-ionization lines, including [Ca~{\sc iv}] $3.207$-{\textmu}m, [Ca~{\sc v}] $4.159$-{\textmu}m, [Ar~{\sc v}] $7.902$-{\textmu}m, [Ar~{\sc iii}] $8.991$-{\textmu}m, [S~{\sc iv}] $10.51$-{\textmu}m,  [Ar~{\sc v}] $13.10$-{\textmu}m,
[S~{\sc iii}] $18.71$-{\textmu}m, [Ar~{\sc iii}] $21.83$-{\textmu}m, [O~{\sc iv}] $25.89$-{\textmu}m and [Fe~{\sc ii}] $25.99$-{\textmu}m (Fig. \ref{fig:photcalc}).   
Most of these lines have previously been observed in young, oxygen-rich SNRs (supplementary text). 
However, in SN 1987A the lines at wavelengths higher than
$\sim 8$ {\textmu}m   are much weaker than predicted (relative to the [Ar~{\sc ii}] line) or absent entirely, although we do detect faint emission for [S~{\sc iv}] $10.51$-{\textmu}m, [S~{\sc iii}] $18.71$-{\textmu}m
(fig. \ref{fig:line_prof_SIII_SIV_FeII_SM}).

\section*{Dust effects and the observed emission}

Infrared observations 
have shown that a large amount of dust was formed in the ejecta of SN1987A  \cite{Wooden1993,Matsuura2011}, consisting of a mixture of silicates and graphite grains \cite{Matsuura2015, Wesson2015}.  However, the composition of the dust close to the center is likely to differ from that of the outer regions \cite{MM}.
Dust nucleation models have predicted that  silicates  dominate over carbon-rich dust in the central regions,   
\cite{Sarangi2015,Sarangi2018,Sluder2018,Sarangi2022}, of which  
Mg$_2$SiO$_4$ (forsterite) is expected to have the highest abundance. Observations of dust in young SNRs have shown that silicates are the  
dominant type of dust \cite{MM}.

Silicate dust grains strongly absorb wavelengths longer than
$\sim 8$~{\textmu}m (fig. \ref{fig:sil_labs})).
This is a potential explanation for why we do not detect the long-wavelength lines at the luminosities predicted by the photoionization models.

We are unable to estimate the optical depth $\tau_{\rm abs}$
at those wavelengths, because of the unknown
composition, grain size, and clumping. However, $\tau_{\rm abs}$ is likely to be large \cite{Cigan2019}. Figure \ref{fig:photcalc}  also shows the dust absorption expected for an assumed optical depth $\tau_{\rm abs}= 6.5$ at $10$~{\textmu}m and amorphous forsterite composition \cite{Gail2020} and the line luminosities after including this dust absorption.
Althopugh the predicted luminosities of lines shorter than  $8$~{\textmu}m are only slightly affected, those at longer wavelengths are substantially suppressed. For [S~{\sc iv}] $10.51$-{\textmu}m and [S~{\sc iii}] $18.71$-{\textmu}m lines, this correction brings the model predictions close to  the observed line luminosities. The observed upper limit on  the [Ar~{\sc iii}] $8.991$-{\textmu}m line is consistent with the model.

The CNS model (Fig. \ref{fig:photcalc}A) predicts strong [Si~{\sc i}] lines at  1.607 and 1.645 ~{\textmu}m. These are not detected in our observations, perhaps because they are blended with broad lines from the ejecta. Dust scattering, dust depletion and ionization of silicon by the ultraviolet flux from the ejecta and ring could all reduce the predicted luminosities  of these lines \cite{MM}. 

We have also modeled the expected line luminosities from a PWN shock \cite{MM}. Most of the observed lines are well reproduced (supplementary text) (fig. \ref{fig:pwn_shock}), without requiring dust effects. However, the   [S~{\sc iii}] $18.71 $-{\textmu}m line is underproduced by a factor $\sim 25$. 

\section*{The nature of the compact object}

Given the uncertain model assumptions about the ionizing spectrum, density, elemental abundances and dust properties, we conclude that the observed lines can be explained by either the CNS or PWN photoionization models. Both require the presence of a NS. Because thermal emission from the CNS should always be present at some level, a combination of both models is also possible.  For the model parameters adopted above, the PWN shock model is less likely but not  excluded \cite{MM}.

There have been previous indications of emission from a central object in SN 1987A. 
Sub-millimeter dust maps of the ejecta show  
a peak in the temperature close to the central region, which might be due to extra energy input from a central source \cite{Cigan2019}. The peak temperature position does not coincide with the location of maximum [Ar~{\sc vi}] emission in our observations \cite{MM}; however, this does not rule out additional heating from a compact object \cite{Page2020}. 
X-ray observations of SN 1987A have detected nonthermal emission at energies above 10 keV, which has been interpreted as being due to a central source \cite{Greco2021,Greco2022}; this would support the PWN scenario. However, the same observations have alternatively been interpreted as thermal emission from the shocks in the ER \cite{Alp2021}.

The 
position and velocity of the argon emission lines constrain any natal NS kick velocity - a velocity imparted on the NS during its formation by anisotropies in the explosion mechanism. 
The offset in position of the narrow emission components corresponds to a velocity $416 \pm 206 \kms$ given the time since the explosion (supplementary text).
However, converting this to a kick velocity 
depends on the separation between the NS and the line-emitting region, which differs between the CNS and PWN models. 

Our observations of high ionization emission lines in the centre of the SN 1987A remnant indicate the presence of a source of ionizing photons,  probably a NS. We cannot determine which scenario is more likely -- young CNS or PWN -- but both require the presence of a NS. 

\begin{figure}
\centering \includegraphics[width=0.6\columnwidth,angle=0]{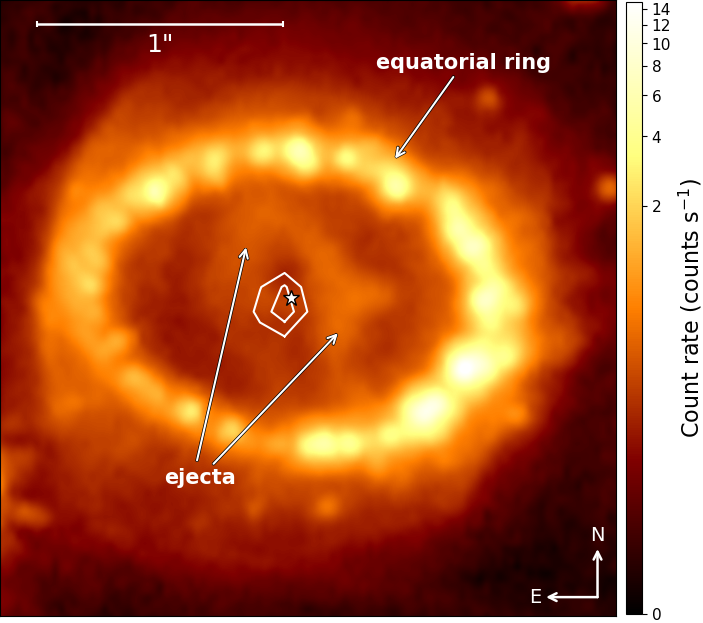}
    \caption{ {\bf Optical image of SN~1987A taken in 2022, 35 years after explosion. }
     Hubble Space Telescope (HST) in the F625W filter \cite{MM}, which is dominated by H$\alpha$ emission. The freely expanding inner ejecta and ER are labelled. White contours mark the [Ar~{\sc vi}] 4.529-{\textmu}m emission observed with NIRSpec (40 and 70\% of the maximal surface brightness in Fig.~\ref{fig:ArII_map}S). The white star denotes the center of the ER \cite{Alp2018}. 
    }
    \label{fig:components}
\end{figure}

\begin{figure}
\centering
\includegraphics[width=13cm]{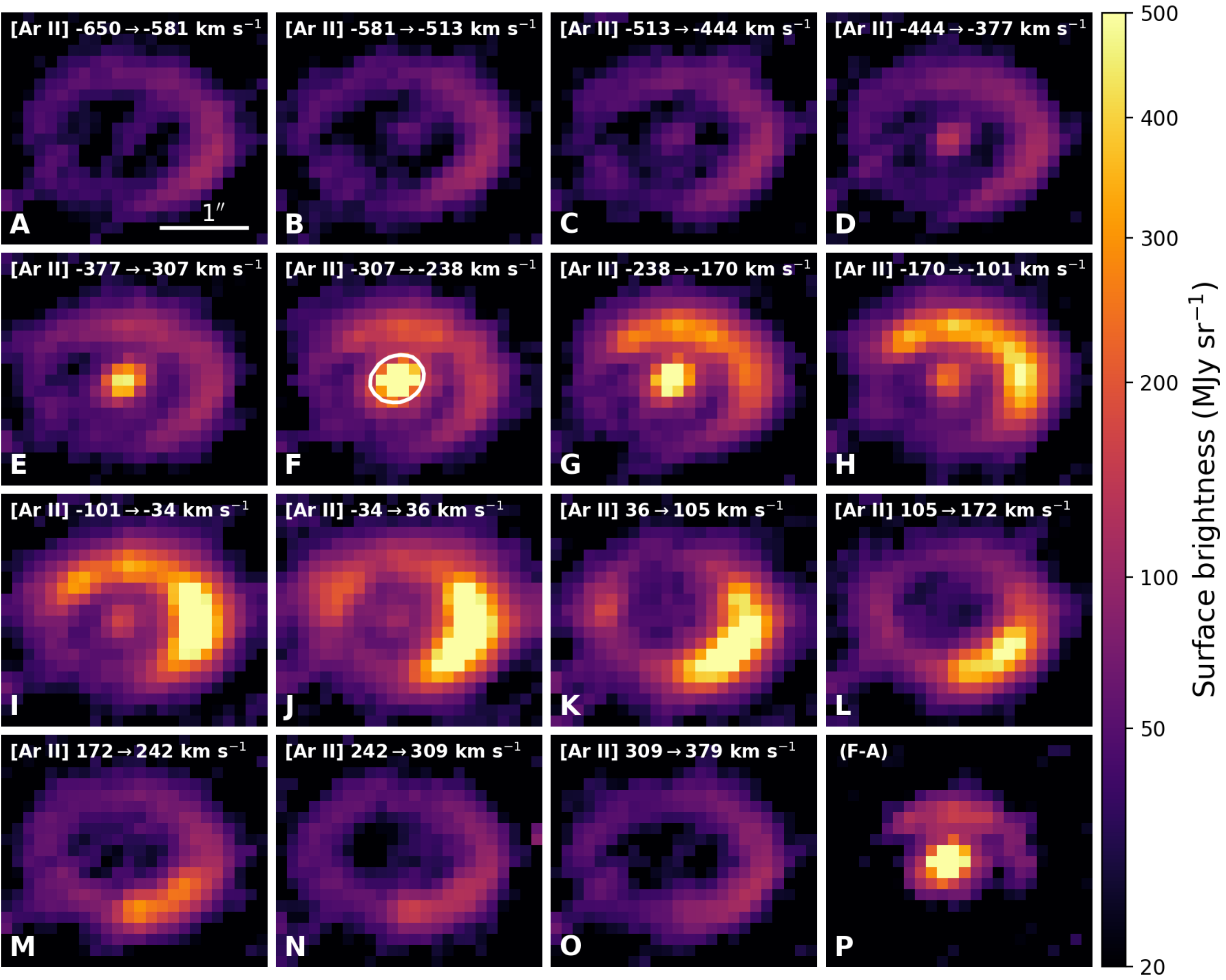}  
\includegraphics[width=13cm]{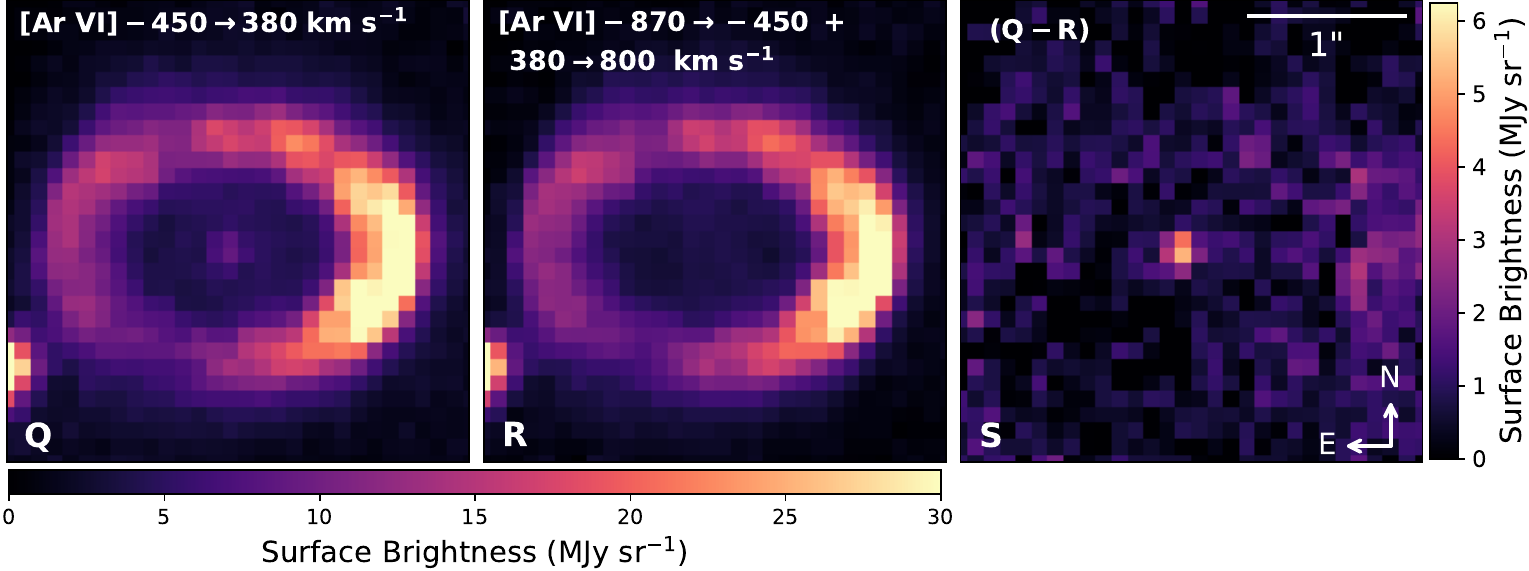}
\caption{\textbf{JWST observations of the [Ar~\textsc{\textbf{ii}}]  and [Ar~\textsc{\textbf{vi}}] lines in SN~1987A.} (\textbf{A} to \textbf{O}) Velocity slices of the MIRI/MRS data cube around the [Ar~{\sc ii}] line at $6.985$~{\textmu}m. Labels in each panel indicate the corresponding velocity range, measured with respect to the systemic velocity of SN~1987A \cite{Groningsson2008}.
Emission is apparent from a central source, which has its maximum intensity offset in velocity from the ER by $\sim -250 \kms$. The white circle in (F) indicates the region used to extract the spectrum in Fig.~\ref{fig:line_prof}A. (\textbf{P}) Residuals between (F) and (A).
 (\textbf{Q}) NIRSpec observations of the [Ar~{\sc vi}] $4.529$-{\textmu}m line, combining all velocity slices between $-450$ and  $+380 \kms$. (\textbf{R}) The NIRSpec data stacked outside that velocity range. (S) Residuals between panels (Q) and (R), separating the emission from the central source. The bright point in the lower left areas of (Q) and (R) is an unrelated star. Color bars are in units of megajansky (MJy) per steradian. 
}
\label{fig:ArII_map}
\end{figure}

\begin{figure}
\centering    
\includegraphics[width=0.9\columnwidth]{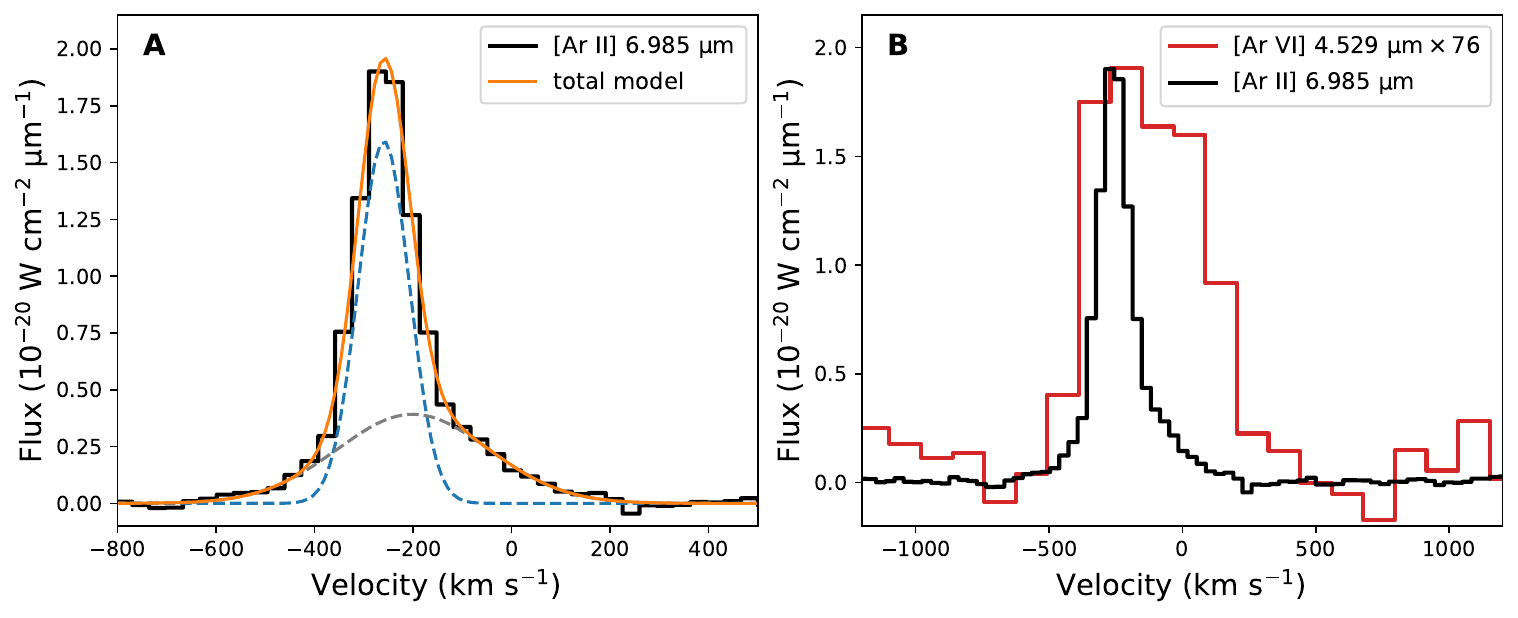}
\caption{\textbf{Velocity profiles of the [Ar~\textsc{\textbf{ii}}] 6.985-{\textmu}m and [Ar~\textsc{\textbf{vi}}] 4.529-{\textmu}m lines from the central source.} (\textbf{A}) The [Ar~{\sc ii}] line in the MRS data (black histogram), extracted from the region indicated in Fig.~\ref{fig:ArII_map}F. The orange curve is a model fitted to the data, consisting of two Gaussian components (blue and grey dashed curves).
(\textbf{B}) The same [Ar~{\sc ii}] data (black histogram), compared to the [Ar~{\sc vi}]  line from the NIRSpec data (red histogram). The [Ar~{\sc vi}] data have a lower spectral resolution and have been scaled by a factor of 76 for display.}
\label{fig:line_prof}
\end{figure}

\begin{figure*}
\begin{center} 
\includegraphics[width=12.cm]{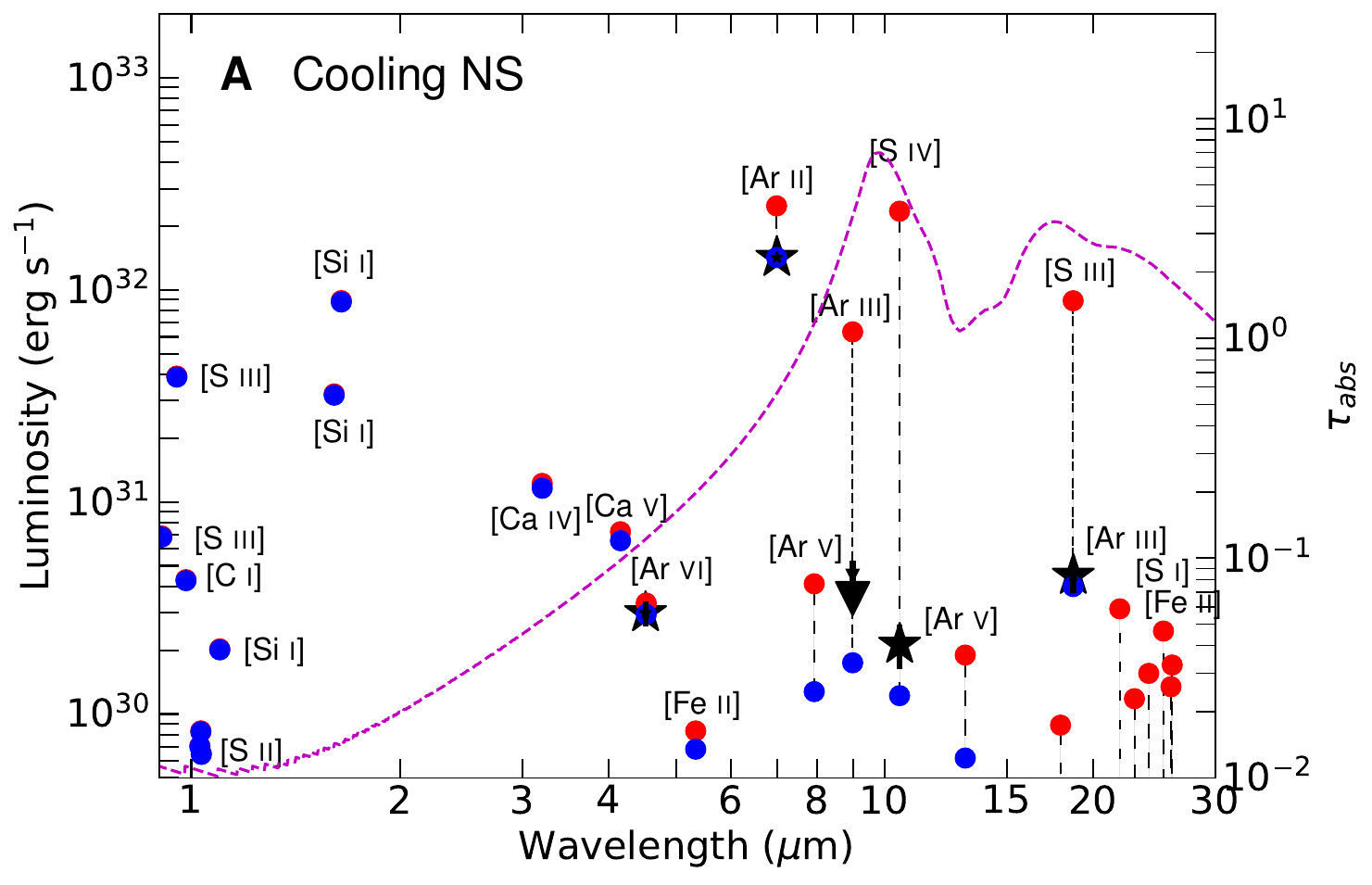}
\includegraphics[width=12.cm]{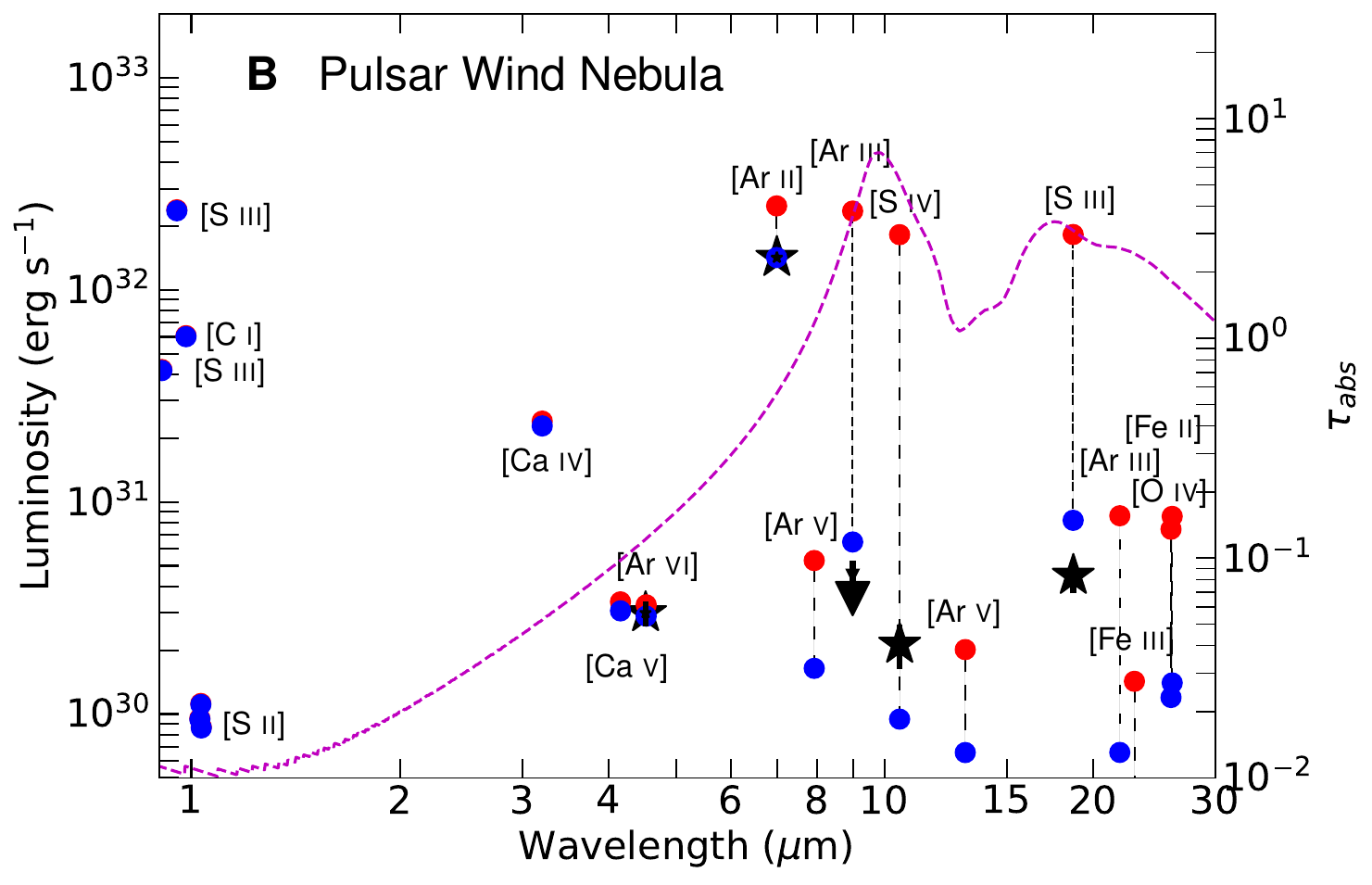}
\end{center}
\caption{\textbf{Photoionization models compared to observed lines.} (\textbf{A}) Model of the oxygen-burning layers expected in a SN
ionized by a young CNS. (\textbf{B}) Model of the same material ionized by  
a PWN.
The optical depth due to silicate dust (magenta curve, on the right axis), assumes Mg$_2$SiO$_4$ composition and $\tau_{\rm abs}=6.5$ at 10~{\textmu}m in both cases. Circles show the model-predicted line luminosities (left axis)before (red) and after (blue) including dust absorption, which are connected by dashed vertical lines. Line identifications are labelled near each red circle. Black stars are the  observed luminosities (Table \ref{table:1}) for the [S~{\sc iii}], [S~{\sc iv}], [Ar~{\sc ii}] and [Ar~{\sc vi}] lines, with error bars showing 1$\sigma$ uncertainties, which are mostly smaller than the symbol size. The upper limit for the [Ar~{\sc iii}] line (black triangle) is 3$\sigma$.
\label{fig:photcalc}}
\end{figure*}
\clearpage
\newpage
\bibliography{scibib}
\bibliographystyle{Science}

\newpage

\textbf{ACKNOWLEDGEMENTS:}
We thank R. Chevalier, T. Janka, S. Woosley and the anonymous referees for constructive comments on the paper and useful discussions.  
MIRI development was supported by NASA, the European Space Agency (ESA), the Belgian Science Policy Office (BELSPO), the Centre Nationale d’Etudes Spatiales (CNES), the Danish National Space Centre, Deutsches Zentrum fur Luft-und Raumfahrt (DLR), Enterprise Ireland, Ministerio De Economiá y Competividad, Netherlands Research School for Astronomy (NOVA), Netherlands Organisation for Scientific Research (NWO), Science and Technology Facilities Council, Swiss Space Office, Swedish National Space Agency, and UK Space Agency
This study is based on observations made with the NASA/ESA/CSA JWST. The data were obtained from the Mikulski Archive for Space Telescopes at the Space Telescope Science Institute, which is operated by the Association of Universities for Research in 
Astronomy, Inc., under NASA contract NAS 5-03127 for JWST.

\textbf{Funding:}
C.F., J.L. and G.Ö. acknowledge support from the Swedish National Space Agency. J.L. acknowledges support from the Knut \& Alice Wallenberg Foundation. 
M.J.B. acknowledges support from European Research Council Advanced Grant ERC-2015-AdG-694520-SNDUST.
P.J.K. and J.J. acknowledge support from Science Foundation Ireland/Irish Research Council Pathway programme under Grant Number 21/PATH-S/9360.
O.C.J. acknowledges support from an STFC Webb fellowship. 
M.M. and N.H. acknowledge support through NASA/JWST grant 80NSSC22K0025. M.M. and L.L. acknowledge support from the NSF through grant 2054178.
J.H. was supported by a VILLUM FONDEN Investigator grant (project number 16599).
O.N. acknowledges support from STScI Director's Discretionary Fund.
T.Tikkanen acknowledges financial support from the UK Science and Technology Facilities Council, and the UK Space Agency.  Part of the research of M.M. and N.H. was carried out at the Jet Propulsion Laboratory, California Institute of Technology, under a contract with the National Aeronautics and Space Administration (80NM0018D0004).
L.C. acknowledges support by grant PIB2021-127718NB-100 from the Spanish Ministry of Science and Innovation/State Agency of Research
MCIN/AEI/10.13039/501100011033. J.A.D.L.B. and B.V. thank the Belgian Federal Science Policy Office (BELSPO) for the provision of financial support in the framework of the PRODEX Programme of the European Space Agency (ESA).

\textbf{Authors contributions:} 
C.F. led the project, performed the modeling, wrote most of the text and contributed to the data analysis.  
M.J.B. and P.B. wrote the proposal science case, designed the observations and coordinated the SN~1987A Guaranteed Time Observations (GTO) Program \#1232. M.J.B. also contributed to the data analysis and writing.
P.J.K. performed the MIRI MRS data reduction and contributed to the analysis and writing.
J.L.  contributed to the data analysis and writing.
O.C.J. designed the observations and was the science lead for the SN~1987A observations program. 
B.S. performed the NIRSpec data reduction. 
M.M. is the co-PI of the SN~1987A observations program, provided three of the nine hours of GTO time used and contributed to the data analysis.
T.Te. contributed to the interpretation of the data.
A.S.H. optimized the observational program. N.H. and A.R. assisted with data processing.
A.C., A.G., J.A.D.L.B., J.H., J.J., L.L., O.K., O.N., O.D.F., R.G., R.M.L., R.W. and T.Ti. reviewed the manuscript and provided comments for incorporation. G.S.W., L.C., E.F.v.D., M.G., Th.H., P.-OL., G.\"O., T.P.R., B.V.~are MIRI PI and co-PIs, who reviewed the manuscript and provided comments for incorporation.

\textbf{Competing interests:} The authors declare that they have no competing interests. 

\textbf{Data and materials availability:} 
The observations are available at the Mikulski Archive for Space Telescopes https://mast.stsci.edu/ under proposal IDs 1232 and 1524 for JWST, and 16789 for HST. The specific MRS and NIRSpec SN1987A observations we used are archived \cite{MRS,NIRSPEC}, as are the 10 Lac observation used for calibration \cite{10LAC} and the HST observations \cite{HST}.

The source code used for the PWN and CNS models is available at 
https://github.com/claesob/SN-codes and archived at Zenodo \cite{PhotoionCode}, while the source code for the shock models is available at 
https://github.com/claesob/SN87A\_shock and archived at Zenodo \cite{ShockCode}. 

\textbf{Supplementary Materials}
\renewcommand{\thefigure}{S\arabic{figure}}
\setcounter{figure}{0}

\renewcommand{\theequation}{S\arabic{equation}}

\renewcommand{\thetable}{S\arabic{table}}
\setcounter{table}{0}

\begin{itemize}
    \item [] Materials and Methods
    \item [] Supplementary Text
    \item [] Figs. S1 to S12
    \item [] Tables S1 to S3
    \item [] References (58-125)
\end{itemize}

\newpage

\setcounter{page}{1}

\begin{center}

\begin{figure}
\centering    
\includegraphics[width=0.3\columnwidth]{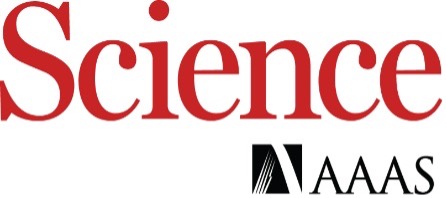}
\end{figure}
\centering
\vspace*{0.5cm}
{\Large Supplementary Materials for} \\
\vspace*{0.5cm}
\textbf{\large {Emission lines due to ionizing radiation from a compact object in the remnant of Supernova 1987A} }
\vspace*{0.5cm}

{C. Fransson${}^*$, M.\ J.\ Barlow, P.\ J.\ Kavanagh, J. Larsson,
O.\ C.\ Jones, B.\ Sargent,   M.\ Meixner, P.\ Bouchet,
T.\ Temim,
G.\ S.\ Wright,
J. A. D. L. Blommaert, N.\ Habel , A.\ S.\ Hirschauer, J. \ Hjorth,   
L.\ Lenki\'{c},   
T.\ Tikkanen, R.\ Wesson,  A. \ Coulais, O.\ D.\ Fox, R. \ Gastaud, A.\ Glasse,  J. Jaspers, O. Krause, R.\ M. Lau, O. Nayak, A. \ Rest, 
L. Colina, \ E.\ F.\ van Dishoeck, M. \ G\"udel Th. \ Henning, P.-O.\ Lagage,  G.\ \"Ostlin,  T. \ P.\ Ray, B.\ Vandenbussche }

${}^*$ Corresponding author: claes@astro.su.se

\vspace*{1cm}
\end{center}
 
\noindent {\bf This PDF file includes:}\\

\noindent
\hspace{1.1cm} Materials and Methods\\
\noindent\hspace*{1.1cm} Supplementary Text\\
\noindent\hspace*{1.1cm} Figs. S1 -- S12\\
\noindent\hspace*{1.1cm} Tables S1 -- S3\\
\noindent\hspace*{1.1cm} References 58 -- 125\\

\newpage

\section{Materials and methods}

\subsection{Analysis of the narrow emission lines from the ejecta}
\label{sec:lines_SM}

Detailed descriptions of the MIRI/MRS and NIRSpec/IFU observations (Proposal ID 1232, Principal Investigator: G. Wright) and data reduction can be found in sect.~2 of \cite{Jones_overview_2023} and sect.~2 of \cite{Larsson2023}, respectively. 
The MRS observations were obtained on 2022 July 16 and the IFUs provided complete spectral coverage from 4.9 to 27.9 {\textmu}m at medium spectral resolving powers $R = \lambda/\Delta\lambda \sim$ 4000--1500 \cite{Jones2023}. Here $\Delta\lambda$ is the FWHM of of a spectral line at wavelength $\lambda$. The MRS fields of view range from $3.''2 \times 3.''7$ in channel 1, the  shortest wavelength channel, to $6.''6 \times 7.''7$ in the long wavelength channel 4 \cite{Wells2015}. The total MRS integration time was 9396~s, corresponding to approximately 3.75 hr when including all overheads. The NIRSpec/IFU observations were obtained on the same date, using the G140M/F100LP, G235M/F170LP, and G395M/F290LP grating/filter combinations. The gratings covered the wavelength ranges 0.97--1.88~{\textmu}m (G140M), 1.66--3.15~{\textmu}m (G235M), and 2.87--5.20~{\textmu}m (G235M), with a spectral resolving power that increased from $R\sim 700$ at the shortest wavelengths to $R \sim1300$ at the longest wavelengths in each grating \cite{Jakobsen2022}. The total exposure times were 1751~s for each of G140M and G235M, and  1225~s for G395M. The NIRSpec/IFU field of view was $3.''1 \times 3.''2$ at all wavelengths. The fields of view of both instruments covered both the ER and the ejecta.  

We used a recent HST Wide Field Camera 3 (WFC3) observation in the F657N filter (obtained on 2022 September 6, program ID 16789) to perform the alignment between the MRS and NIRSpec data. Another HST WFC3 image from this program taken in the F625W filter (obtained on 2022 September 5) was used for illustrating the main emission components of the system in Fig.~\ref{fig:components}. Both observations consisted of four dithered exposures, with total exposure times of 2600~s (F657N) and 1080~s (F625W).  The exposures were combined using DrizzlePac \cite{Gonzaga2012}, which also applies distortion corrections and removes cosmic rays. The resulting images were subsequently aligned with Gaia data release 3 \cite{Gaia2023} to improve the astrometry.  

\subsubsection{Spatial distribution of the Ar emission}
\label{sec:spatial_ar}
Astrometric uncertainties can be introduced into JWST data by guide star catalogue errors and roll uncertainty \cite{Pont2022}. Therefore, the MRS and NIRSpec data had to be spatially aligned to extract spectra from identical regions and compare the spatial distribution of the emission. For this we used the HST WFC3 obseervations above. 
 
For NIRSpec, two stars within the field of view (FOV), as well as several hotspots in the ER, were used to perform the alignment with HST \cite{Larsson2023}. However, no suitable stars were available within the FOV of MRS. The ER of SN~1987A was therefore used to perform the alignment for MRS. We produced a continuum-subtracted image of the H~{\sc i} 7.460~{\textmu}m line, which we expect to originate from the same hotspots in the ER as the H$\alpha$ emission in the HST F657N image \cite{Larsson2019}. The limited spatial resolution of the MRS means that most of the hotspots are blended into a diffuse structure. We fitted three bright hotspots with Gaussians, from which we determined the offset of their centroids from the corresponding hotspots in the HST image. In addition, we fitted the whole ER with an elliptical model with diffuse emission (described in \cite{Alp2018}) and compared the offsets of the centers of the ellipses, which gave consistent results for the offset. The uncertainty in the alignment between MRS and HST estimated from the hotspot fits was  $\sim 0.''03$, which dominates the uncertainty in the comparison between MRS and NIRSpec. 

To assess the spatial distribution of the [Ar~{\sc ii}] narrow component and whether it is spatially resolved, we compared the radial profile to an empirical MRS PSF. While Star~3 (RA=05:35:28.26, Dec=-69:16:11.89) is seen in the MRS spectral band 1C datacube, its location of at the very edge makes it unsuitable for such a comparison. Instead we made use of a JWST calibration observation of a spatially unresolved source, the O-star 10~Lac (RA=22:39:15.67, Dec=+39:03:00.97). We reprocessed these data (PID~1524, PI: D.~Law) following the same method described in \cite{Jones_overview_2023} to produce an MRS band 1C datacube. To enable a direct comparison between the two, we constructed datacubes for both 10 Lac and SN1987A in the MRS IFU coordinate system so the broadening in the along-slice direction is in the same dimension \cite{Argyriou2023}. 

\begin{figure}
    \centering    \includegraphics[width=0.45\columnwidth,angle=0]{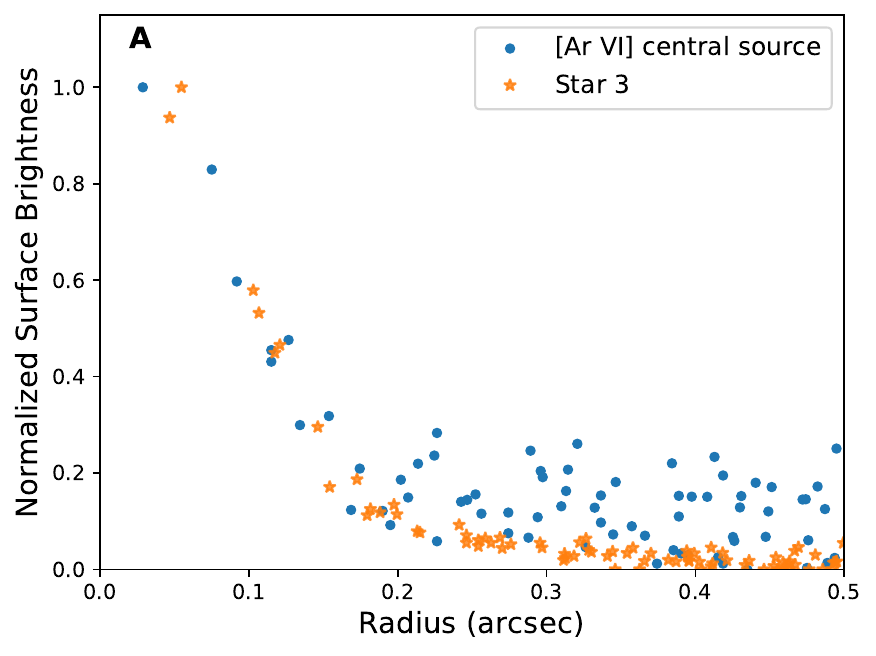}
    \centering    \includegraphics[width=0.45\columnwidth,angle=0]{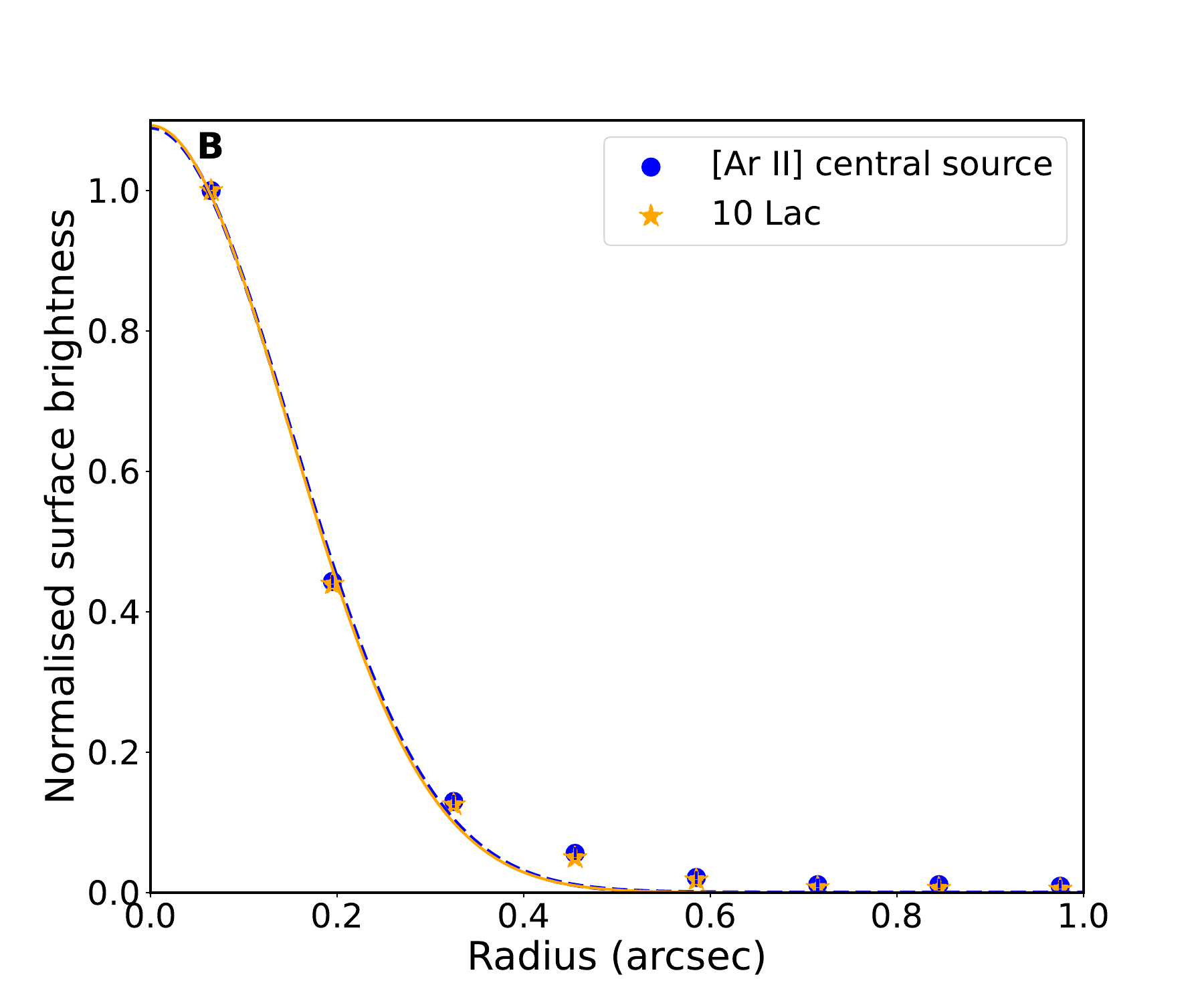} 
\caption{ \textbf{Radial profiles of the [Ar~\textsc{\textbf{vi}}] and [Ar~\textsc{\textbf{ii}}] lines.} (A) Profile of the [Ar~{\sc vi}] line (blue circles) and `Star 3' (orange stars). The enhanced emission in the [Ar~{\sc vi}] profile compared to Star 3 outside $0.''2$ may be due to residual continuum emission from the ejecta. (B) Profile of the narrow component of the [Ar~{\sc ii}] line (blue circles). The flux profile of 10~Lac (orange stars) is included as a comparison of an unresolved points source in MRS band 1C. The best-fit Gaussian models are shown by the overlapping dashed blue and solid orange lines for the [Ar~{\sc ii}] source and 10~Lac, respectively.}
\label{fig:ArVI_extraction_SM}
\end{figure}

From the 10 Lac data, we isolated and summed the velocity slices around the [Ar~{\sc ii}] line and determined the source centroid by fitting a two-dimensional Gaussian model to the PSF peak. We defined annular bins of 1 pixel width centered on the centroid and determined the surface brightness in each bin before normalising by the innermost bin. The resulting 10 Lac radial profile is shown in Fig.~\ref{fig:ArVI_extraction_SM}B.

To extract the radial profile of the [Ar~{\sc ii}] narrow component we isolated the narrow component from the [Ar~{\sc ii}] broad component, the [Ar~{\sc ii}] ER emission, and the dust continuum in the ring. This was performed by first producing a datacube in a narrow velocity range around the [Ar~{\sc ii}] line. Datacubes were also produced from narrow velocity ranges to the low and high end of the [Ar~{\sc ii}] line and used to measure and subtract the continuum in the [Ar~{\sc ii}] datacube, removing the dust continuum. We fitted the spectra extracted from each pixel of the continuum-subtracted datacube with a multi-component Gaussian model. The central wavelengths and widths of the two components representing the broad and narrow [Ar~{\sc ii}] line were fixed to the best fitting values in Table.~\ref{table:1}, which were determined from the spectrum extracted from the entire ejecta region. Additional Gaussian components were included to account for the [Ar~{\sc ii}] emission from the ER. The resulting models were subtracted to leave only the [Ar~{\sc ii}] narrow component in the datacube. As with the 10~Lac cubes, the frames were then summed, the centroid determined, and the radial profile extracted using annular bins of 1 pixel width. The resulting radial profile and comparison with 10~Lac is shown in Fig.~\ref{fig:ArVI_extraction_SM}B.  
We fitted the profiles with Gaussian models to determine the FWHMs of the [Ar~{\sc ii}] narrow component and 10~Lac to be $0.''353~(\pm 0.''006)$ and $0.''350~(\pm0.''005)$, respectively. The best fits are shown in Fig.~\ref{fig:ArVI_extraction_SM}-right. We therefore conclude that the [Ar~{\sc ii}] narrow component is spatially unresolved.

We performed an equivalent analysis of the radial profile of the [Ar~{\sc vi}] emission compared with the radial profile of Star 3 (Fig.~\ref{fig:ArVI_extraction_SM}A), which represents an unresolved point source. 
The radial profile of the [Ar~{\sc vi}] source was obtained from the continuum subtracted image (Fig.~\ref{fig:ArII_map}S), while that of the star was obtained from the original image at the same wavelength (Fig.~\ref{fig:ArII_map}Q).  The lower signal and lower spectral resolution in the [Ar~{\sc vi}] line compared to [Ar~{\sc II}] makes the separation of the line emission into two components uncertain, so we show the profile of the full line in Fig.~\ref{fig:ArVI_extraction_SM}A. 
We find that the peak of the emission is unresolved, while there is some evidence of more extended emission outside 0."2. However this could be due to imperfect continuum subtraction.

The centroid of the [Ar~{\sc vi}]  source in the background-subtracted image is located $38\pm22$~milliarcseconds (mas) east and $31\pm22$  mas south of the geometric center of the ER determined from HST optical images \cite{Alp2018}. 
For comparison, the centroid of the blue component of the [Ar~{\sc ii}] emission is located  $65\pm26$~mas east and $51\pm30$ mas south of the geometric center. The [Ar~{\sc ii}] and [Ar~{\sc vi}] positions are consistent with each other and within 3 sigma of the geometric center. 
The uncertainty in the [Ar~{\sc vi}] position is dominated by statistical uncertainties, while that of [Ar~{\sc ii}] is dominated by the systematic astrometric uncertainties (see above).

\subsubsection{Spectral line fitting and search for additional lines}

\begin{figure}[h!]
    \centering       \includegraphics[width=0.5\columnwidth,angle=0]{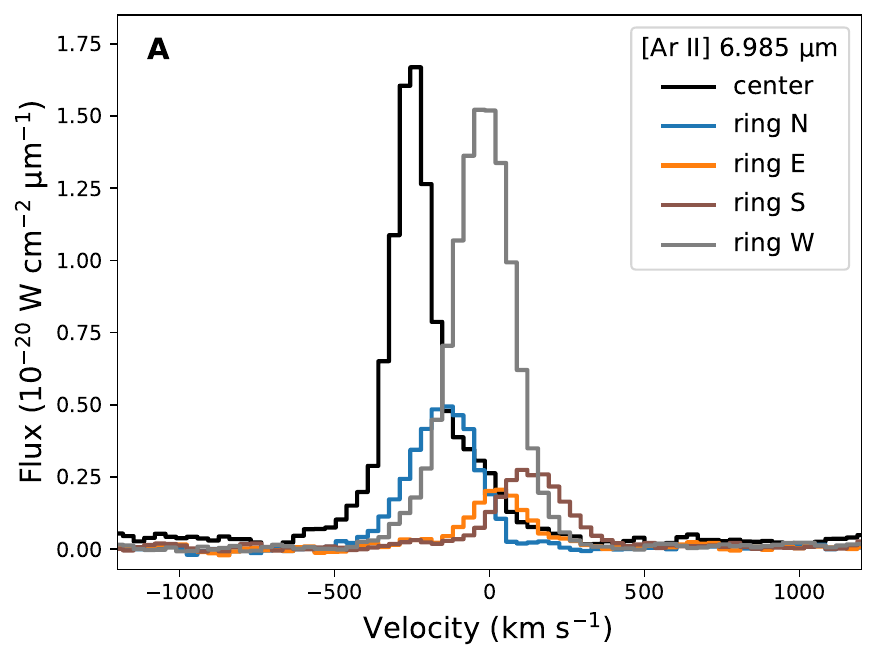}    \includegraphics[width=0.45\columnwidth,angle=0]{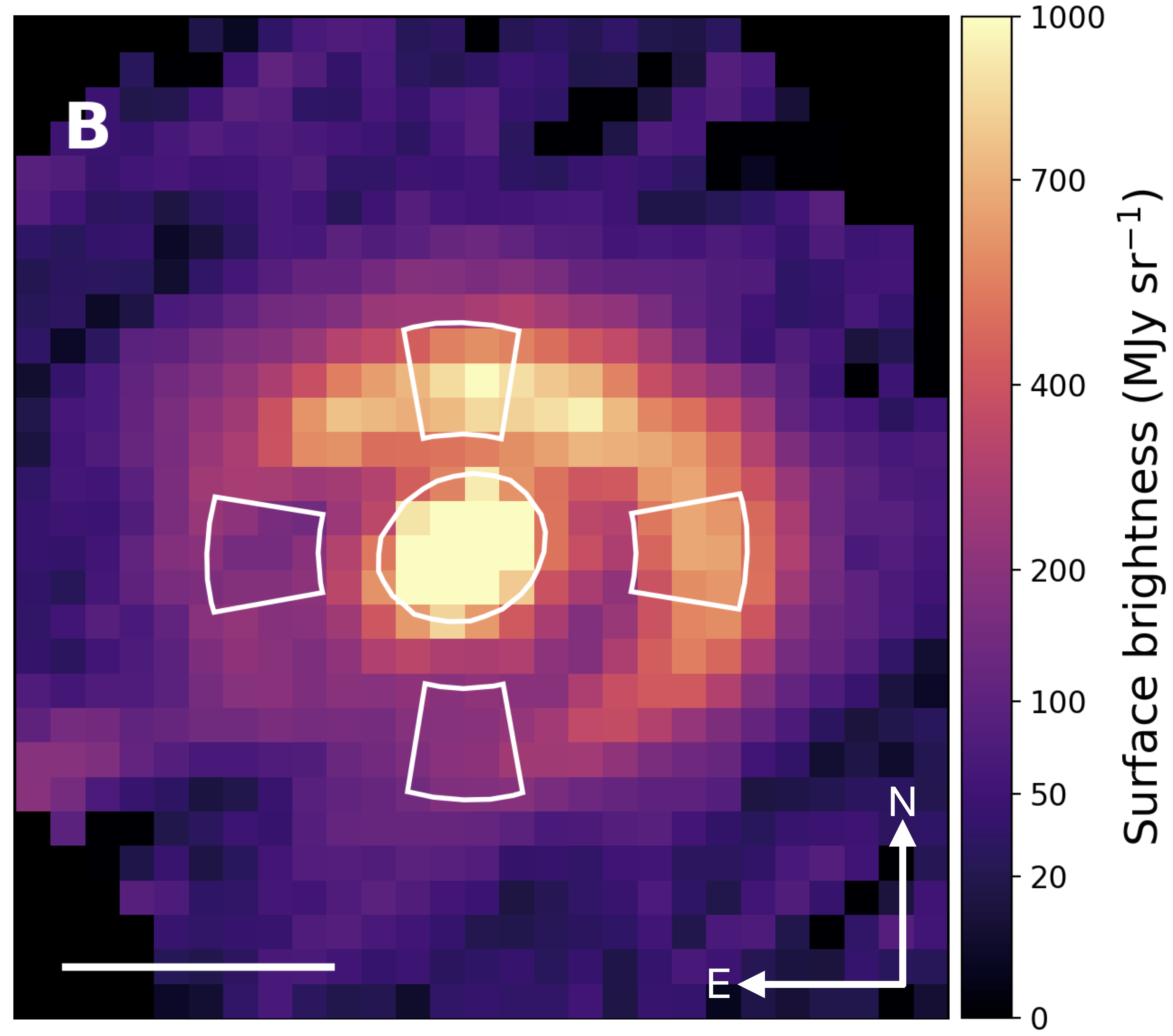}
    \caption{\textbf{Velocity profiles and extraction regions for the [Ar~\textsc{\textbf{ii}}] 6.985~{\textmu}m line.} (A) Velocity profile of the central source (black histogram) together with velocity profiles extracted from different parts of the ER (blue, orange, brown and grey histograms were extracted from the north, east, south and west ER, respectively). (B) Extraction regions for the velocity profiles in A, superposed on an integrated image of the [Ar~{\sc ii}] emission. There is a velocity shift between the central component and the different ER extractions. The offset of the stronger ejecta component of the [Ar~{\sc ii}] line is $-259.6 \pm 0.4 \kms$ and the FWHM is $121.9 \pm 1.3 \kms$, narrower than the FWHM of the ring extractions (see Table~\ref{table:ar2_er}). 
    There is a red extension of the line from the central source.
    }      
\label{fig:line_prof_ArII_ArIV_ER}
\end{figure}

\begin{figure}[h!]
    \centering    \includegraphics[width=0.45\columnwidth,angle=0]{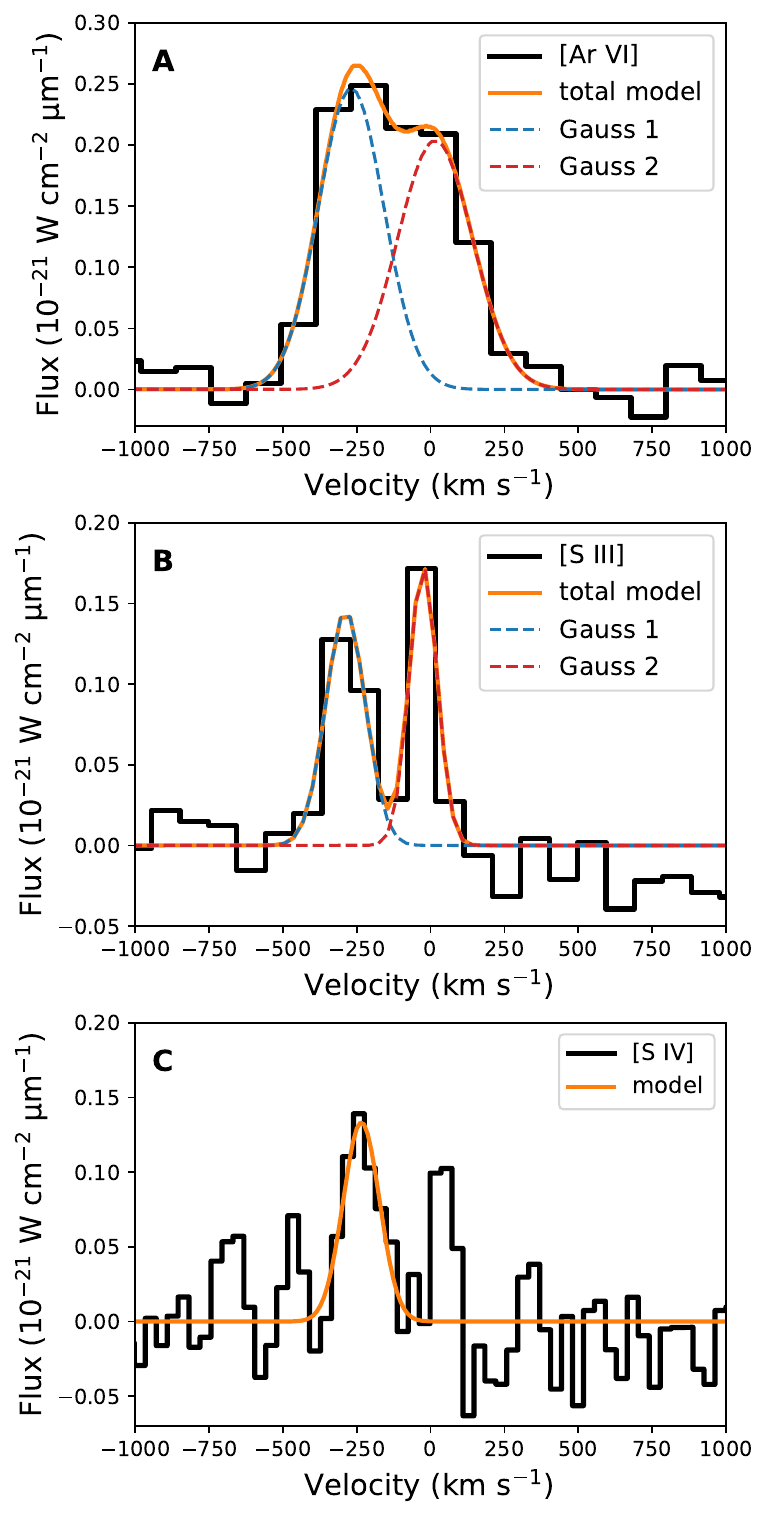}
    \caption{\textbf{Velocity profiles of the [Ar~\textsc{\textbf{vi}}]\ 4.529~{\textmu}m, [S~\textsc{\textbf{iii}}]\ 18.71~{\textmu}m and  [S~\textsc{\textbf{iv}}]\ 10.51~{\textmu}m lines from the central region.} (A) The [Ar~{\sc vi}] line (black histogram). The orange curve is a model fitted to the data, consisting of two Gaussians (blue and red dashed lines). (B) The [S~{\sc iii}] line (black histogram). The orange curve is a model fitted to the data, consisting of two Gaussians (blue and red dashed lines). (C) The [S~{\sc iv}] line (black histogram) fitted by a Gaussian model (orange curve). In the case of [Ar~{\sc vi}], the two Gaussian components are associated with the central source, with a minimal contribution from the ER (cf. Fig.~\ref{fig:ar6_ringcomp}). For [S~{\sc iii}], the red component is dominated by scattered light from the ER. For all three lines, the blue component has a similar velocity.
    }  \label{fig:line_prof_SIII_SIV_FeII_SM}
\end{figure}

To model the emission lines in the MRS spectra we used the Gaussian emission line fitting (\texttt{elf}) routines within the {\sc dipso} spectral analysis package \cite{Howarth2014}. 
For each line component, \texttt{elf} returns an estimate for the best-fitting central wavelength or velocity, simultaneously
with 1$\sigma$ uncertainties on each line's FWHM and total flux
The resulting models are shown in
Figs.~\ref{fig:line_prof}, \ref{fig:line_prof_ArII_ArIV_ER} and \ref{fig:line_prof_SIII_SIV_FeII_SM}.

\begin{table}[h!]
\centering
\caption{\textbf{Parameters for the [Ar~\textsc{\textbf{ii}}] line in different parts of the ER}. The four extraction regions are shown in Fig.~\ref{fig:line_prof_ArII_ArIV_ER}B. The lines were fitted with a Gaussian model. 
}
\vspace{3mm}
\footnotesize{
\begin{tabular}{|l r r |} 
 \hline
 Region & Offset & FWHM  \\ 
& $\kms$ & $\kms$   \\
 \hline\hline
 & & \\ 
\vspace{-3mm}
North  & $-153.3 \pm 2.0$ & $253.0 \pm 4.7$\\
 & & \\ 
\vspace{-3mm}
West  & $-26.6 \pm 0.8$ & $225.8 \pm 1.8$\\
 & & \\ 
\vspace{-3mm}
South  & $127.7 \pm 3.2$ & $263.1 \pm 8.1$\\
 & & \\ 
 \vspace{-3mm}
East  & $22.2 \pm 3.4$ & $231.3 \pm 9.6$ \\
 & & \\ 
\hline
\end{tabular}
}
\label{table:ar2_er}
\end{table}

The NIRSpec spectra were fitted using the \texttt{curve\_fit} function of \textsc{scipy.optimize} \cite{Virtanen2020}, which gave consistent results with {\tt elf}.
The models were constrained by requiring the FWHM of the Gaussians to be no smaller than the spectral resolution at the relevant wavelengths.
For the [Ar~{\sc vi}] line, which is well modelled with two Gaussians (Fig.~\ref{fig:line_prof_SIII_SIV_FeII_SM}),  the uncertainty ranges on the FWHM of both components reached the resolution limit, so only upper limits on the FWHM are provided in Table \ref{table:1}.

While the [Ar~{\sc ii}] $\wl 6.985$~{\textmu}m,  [Ar~{\sc vi}] $\wl 4.529$~{\textmu}m lines are in regions of low dust continuum emission, the comparison with the mid-infrared (MIR) spectra of young supernova remnants (see below)
led us to search for additional lines at longer wavelengths. The strong background from the dust continuum, as well as the decreasing spatial resolution with wavelength makes it increasingly more difficult to separate lines emitted by the central ejecta from the strong ER and interstellar medium (ISM) emission. However, the velocity offset seen in especially the [Ar~{\sc ii}] $\wl 6.985$~{\textmu}m 
helps in separating narrow ejecta components from the ER and ISM contributions. In this way, we detected 
two more weak lines with similar velocity offsets as the [Ar~{\sc ii}] line.  Figure \ref{fig:line_prof_SIII_SIV_FeII_SM} shows the
[S~{\sc iii}] $18.71$~{\textmu}m and [S~{\sc iv}] $10.51$~{\textmu}m line profiles from the central ejecta. 
The velocity offsets and luminosities are summarized in Table \ref{table:1}.
These velocities are consistent with the offset in the [Ar~{\sc ii}] line, and are therefore likely to originate from the same source. We are only able to place a 3$\sigma$ upper limit on the flux of [Ar~{\sc iii}] $8.991$~{\textmu}m, corresponding to a line luminosity upper limit of $<3.5 \times 10^{30} \ergs$.

\begin{figure}
\centering
    \includegraphics[width=0.5\columnwidth,angle=0]{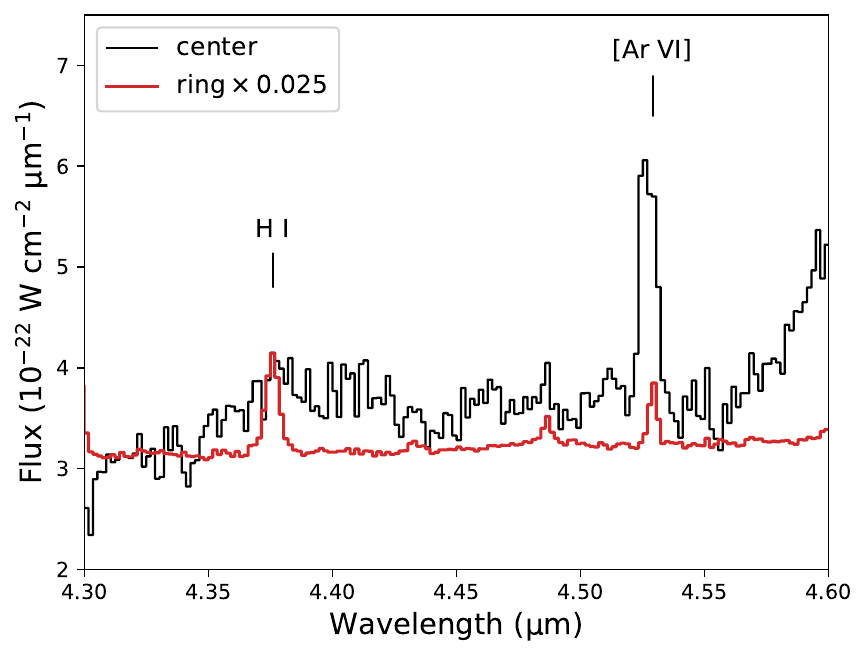}
    \caption{\textbf{Spectra from the central source and the ER at 4.30-4.60~{\textmu}m.} The black and red histograms show spectra extracted from the central region and ER, respectively. A H~{\sc i} line is present at 4.377~{\textmu}m, which is stronger than the [Ar~{\sc vi}] line in the ER, but which does not appear in the central extraction region. This indicates that scattered light from the ER makes a minimal contribution to the [Ar~{\sc vi}] profile extracted from the central region. 
    }
    \label{fig:ar6_ringcomp}
\end{figure}

\begin{figure}
    \centering    
\includegraphics[height=6.cm,angle=0]{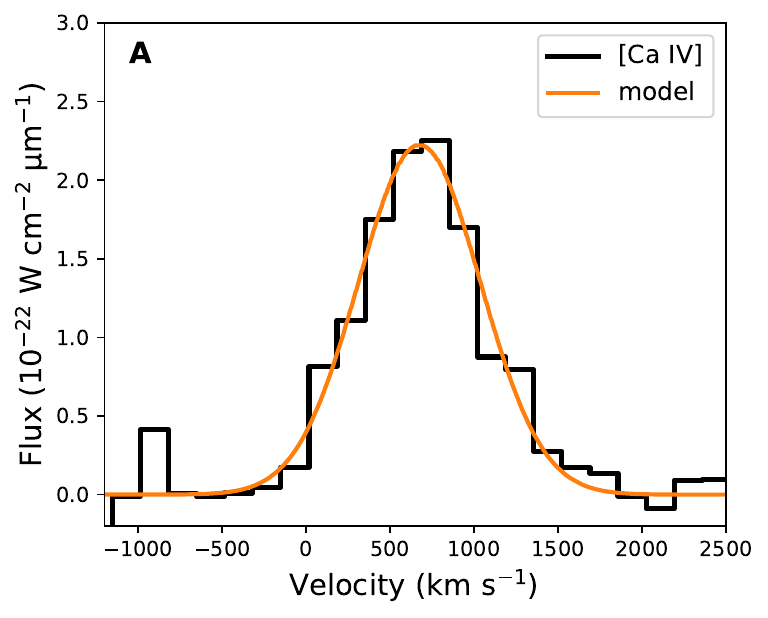}
\includegraphics[height=6.cm,angle=0]{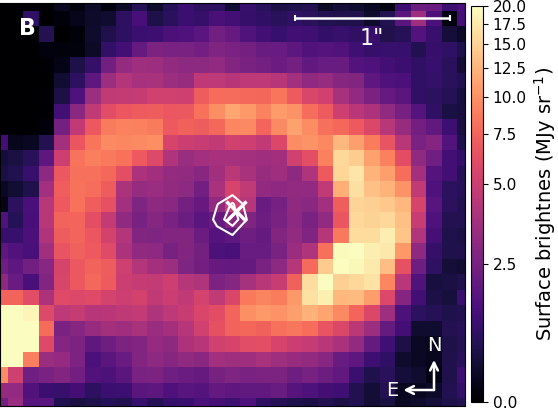}
\caption{\textbf{Line profile and spectral map of the [Ca~\textsc{\textbf{iv}}] 3.207~{\textmu}m line.} (A) Velocity profile extracted from the central region (black histogram) 
fitted with a Gaussian model (orange curve). (B) Velocity slice integrated over the [Ca~{\sc iv}] line in panel A. The white contours show the  [Ar~{\sc vi}] 4.529~{\textmu}m emission (Fig.~\ref{fig:ArII_map}S), while the cross shows the geometric center of the ER \cite{Alp2018}.
    }  \label{fig:line_prof_CaIV_VI_SM}
\end{figure}

\begin{figure}
\centering
    \includegraphics[width=0.5\columnwidth,angle=0]{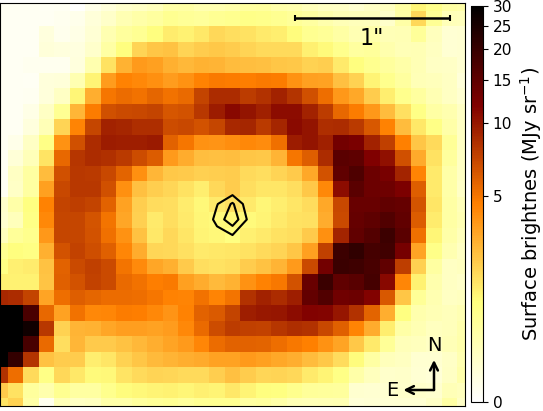}
    \caption{\textbf{Continuum emission in NIRSpec and the central [Ar~\textsc{\textbf{vi}}]  source.} The image of the continuum (color scale) was produced by integrating over the  3.42--3.58~{\textmu}m wavelength interval. The black contours show the central [Ar~{\sc vi}] 4.529~{\textmu}m emission (Fig.~\ref{fig:ArII_map}S).
    There is a weak enhancement in the continuum just north of the [Ar~{\sc vi}] source. 
    }
     \label{fig:continuum}
\end{figure}

In the NIRSpec data we identify two additional lines which could be related to the central source, [Ca~{\sc iv}]  3.207~{\textmu}m and  [Ca~{\sc v}]  4.159~{\textmu}m. The [Ca~{\sc v}] line, however, is likely blended with an H$_2$ line at 4.181~{\textmu}m \cite{Larsson2023}, so its properties are uncertain.  We model the [Ca~{\sc iv}] line with a single Gaussian with FWHM $854 \pm 40 \kms$ and a peak velocity shifted to the red by $676 \pm 17 \kms$ (Fig. \ref{fig:line_prof_CaIV_VI_SM}A), substantially different from the Ar and S lines.
The [Ca~{\sc iv}] emission is located $3\pm10$~mas east and $114\pm15$ mas north of the center (\ref{fig:line_prof_CaIV_VI_SM}B), which implies a significant offset to the north compared to the [Ar~{\sc vi}] emission. 
Expressed in terms of an equivalent velocity in the plane of the sky for the freely expanding ejecta, (with velocity proportional to the distance from the center), the [Ca~{\sc iv}] emission is offset by $757\pm96 \kms$ to the north of the center. We therefore interpret the [Ca~{\sc iv}] line as not directly related to the [Ar~{\sc ii}] emission.   A connection with the compact object is possible, because the emission region and line width are much smaller than observed for the general ejecta, which reaches velocities $\sim 5000 \kms$ close to the ER, while the ionisation state is higher \cite{Larsson2023}. 
The region just north of the center also shows enhanced continuum emission at $3.5$~{\textmu}m (Fig.~\ref{fig:continuum}), whereas there is no point source in the continuum at the location of the Ar emission. 

\subsection{Spectral modeling}
\label{sec:spectral_modeling}
\subsubsection{Photoionization model}
\label{sec:photomodel}

To model the photoionization by the radiation from a PWN or  CNS we update an earlier photoionisation code
\cite{CF1992}. We now include all important ionization states of H, He, C, N, O , Ne, Na, Mg, Al, Si, S, Ar, Ca, and Fe. In \cite{CF1992} Ar was not included in the models, and the Si-S-Ar-Ca zone was not discussed. 

The ionization structure is set (to first order) by the 
the ionization parameter \cite{Tarter1969}, 
\begin{equation}
\xi=\frac{L_{\rm ion}}{n_{\rm ion} r^2}
\end{equation}
which is proportional to the ratio of the number density of ionizing photons to the number density of atoms.  $L_{\rm ion}$ is the ionizing luminosity, $n_{\rm ion}$ the ion number density and $r$ the distance to the source. 
For discriminating between different ionizing sources the spectral shape is also important for the  fractions of the different ionization stages and the temperature, as illustrated by the different models. 

For the ionization equilibrium we include photoionization, collisional ionization, radiative and dielectronic recombination,  including updated recombination data \cite{Badnell2006,Altun2006,Altun2007,Abdel-Naby2012,Bleda2022},   photoionization cross sections were updated from fits in \cite{Verner1996} and the TOP base \cite{TOPbase1993}, and  Auger ionization from the inner shells \cite{Kaastra1993}. Charge transfer between elements heavier than H and He are lacking, with a few exceptions, and introduce an uncertainty.  

The ionizing flux is calculated in the `outward only' approximation, including  diffuse emission from continuum emission and line emission \cite{Kallman1982}, while the radiative transfer in the lines is treated with the Sobolev approximation.  
The line emission is calculated using detailed multilevel atoms in non-local thermodynamical equilibriumnon-local (NLTE), with atomic data from the Chianti Atomic Database \cite{DelZanna2021},  NORAD-ATOMIC-DATA base \cite{Nahar2020}, NIST Atomic Spectra Database \cite{NIST} and the Atomic Line List \cite{vanHoof2018}. 
We adopt updated collision strengths for the important fine-structure lines of [Ar~{\sc ii}]  \cite{Pelan_Ar2coll_1995} and [Ar~{\sc vi}] \cite{Saraph_Ar6_coll_1996},  and for [Ca IV]  \cite{Nahar_Ca4_coll2023}. The calculation is stopped when the temperature reaches 100 K, where no lines below 30 $\mu$m are excited. The updated code is written in Fortran and available on Zenodo \cite{PhotoionCode}, together with the output files.

\subsubsection{Abundances and structure of the core}
\label{sec:ab_struc}
\begin{figure}
    \centering
\includegraphics[width=1.0\columnwidth]{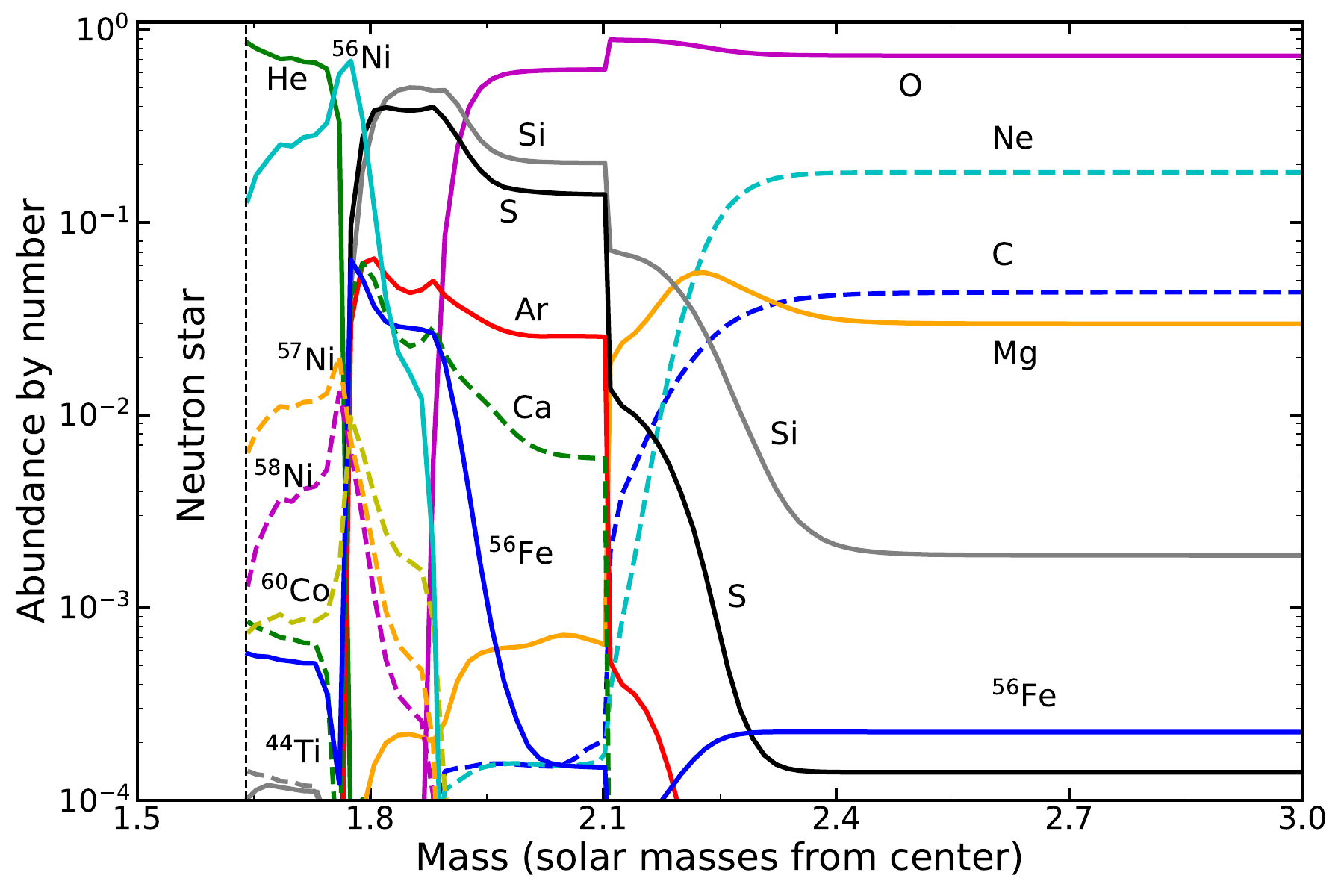}
\caption{\textbf{Fractional abundances by number as a function of mass from the center for the core of a $19 \msun$ model\cite{WoosleyHeger2007}.} Ar is most abundant in the zone between $1.8-2.1 \msun$ from the center. The region interior to $\sim 1.65 \msun$ (dashed vertical line) becomes the NS.
    }
\label{fig:abundances}
\end{figure}

Models of the progenitor and nucleosynthesis of SN 1987A indicate a zero age main sequence (ZAMS) mass of $15-20 \msun$ \cite{McCray2016}. Fig.~\ref{fig:abundances}  shows the predicted composition in the central region of an explosion of a $19 \msun$ ZAMS progenitor \cite{WoosleyHeger2007}, adopted for the modeling. Other models in the 15 to 20 $\msun$ range have similar structure and mainly differ in the varying masses of the different zones. Close to the NS the predicted composition is dominated by He and ${}^{56}$Ni, decaying into ${}^{56}$Fe. 
For the O -- Si -- S -- Ar -- Ca zone between $1.9-2.1 \msun$ in mass coordinates,  
we have taken an average fractional  abundances over the zone (Table \ref{table:abund}). 
Within the zone the abundances of the most abundant elements O, Si, S, Ar and Ca, vary by approximately a factor of two (Fig. \ref{fig:abundances}). For comparison we  model the line emission from from the O -- Ne -- C -- Mg zone in Sect. \ref{sec:O_Ne_C_Mg_models} , with averaged abundances  between $2.10 - 4.0 \msun$ (Table \ref{table:abund}). 
\begin{table}[h!]
\centering
\caption{\textbf{Averaged fractional abundances used in the calculations.}   
}
\vspace{3mm}
\footnotesize{
\begin{tabular}{|l l l |} 
 \hline
Element &\multicolumn{2}{c|}{Abundance by number}\\  
&O--Si--S--Ar--Ca&O--Ne--C--Mg\\  
 \hline\hline
H  & $2.06\times 10^{-7}$&$6.78\times 10^{-9}$\\
He & $2.30\times 10^{-5}$&$1.10\times 10^{-4}$\\
C  & $1.58\times 10^{-4}$&$4.04\times 10^{-2}$\\
N  & $1.92\times 10^{-5}$&$3.42\times 10^{-5}$\\
O  & $5.80\times 10^{-1}$&$7.45\times 10^{-1}$\\
Ne & $1.52\times 10^{-4}$&$1.67\times 10^{-1}$\\
Na & $1.00\times 10^{-9}$&$1.84\times 10^{-6}$\\ 
Mg & $6.46\times 10^{-4}$&$3.12\times 10^{-2}$\\
Al & $1.85\times 10^{-4}$&$2.83\times 10^{-3}$\\
Si & $2.26\times 10^{-1}$&$6.79\times 10^{-3}$\\
S  & $1.56\times 10^{-1}$&$1.17\times 10^{-3}$\\ 
Ar & $2.74\times 10^{-2}$&$1.47\times 10^{-4}$\\
Ca & $8.29\times 10^{-3}$&$3.51\times 10^{-5}$\\ 
Fe & $1.40\times 10^{-3}$&$4.34\times 10^{-4}$\\
\hline
\end{tabular}
}
\label{table:abund}
\end{table}

The density in the SN core is uncertain.The inner ejecta in SN 1987A is both clumpy and anisotropic \cite{Kjaer2010, Larsson2013, Matsuura2017}. To get an estimate of the density we nevertheless approximate this with a constant density sphere. For constant  emissivity the line profile is then parabolic, with flux density $F(V) \propto 1 - (V/V_{\rm core})^2$, where $V$ is the velocity from the line center and $V_{\rm core}$ the maximum velocity of the core. The full width velocity at half maximum, $V_{\rm FWHM}$, is related to this by  $V_{\rm core} = V_{\rm FWHM}/\sqrt 2 $.   

Observations of the  [O~{\sc i}]  $\lambda \lambda $6300 , 6364 lines during the first 500 days after explosion were modeled  with a maximum velocity $\sim 1700 \kms$ \cite{Li1992}.  Similarly, observations of these and other lines in 1995 were modeled with a constant emissivity profile to $\sim 1700 \kms$,  steeply decreasing at higher velocities. A fit with a parabolic line profile to the [O~{\sc i}]  $\lambda $6300 line from 1999 \cite{Fransson2013}  gives a good fit up to $\sim 2200 \kms$, but shows an extension above the parabolic profile at higher velocity, as expected from the 3D observations above.  From this model we find $V_{\rm FWHM}  \approx 3500 \kms$, implying  $V_{\rm core}\approx 2500 \kms$.   A similar fit to the [Fe {\sc i}] $\wl \ 1.443$~{\textmu}m,  [Fe~{\sc ii}] $\wl 1.534$~{\textmu}m,  and  [Fe {\sc ii}] $\wl \ 25.98$~{\textmu}m lines \cite{Jones_overview_2023} give $V_{\rm core}\approx 2700 \kms$. As an average we adopt  $V_{\rm core} = 2200 \pm 500 \kms $.

The ejecta mass inside the oxygen core of a 19 $\msun$ progenitor is $\sim 3.0 \msun$, excluding the NS \cite{WoosleyHeger2007}. 
For a uniform oxygen core density with the above mass and $V_{\rm core} = 2200\kms$,
this corresponds to a density of $\sim 9.66 \times 10^{-20} \ \rm g ~cm^{-3}$ at 35 years, or a number density of ions of $n_{\rm ion} = 2.63 \times 10^{3} (A/22)^{-1} \ \ccm$, where $A $ is the mean atomic weight, $A \sim 30$ in the inner Si zone, $A \sim 22$ in the O -- Si --S --Ar --Ca zone and $A \sim 16.8$ in the O -- Ne -- C -- Mg zone. 

However, both observations and hydrodynamical multidimensional simulations \cite{Gabler2021} have shown that  clumping and filaments produce considerably higher density in the emitting gas. 
The line ratio of the [O~{\sc i}] $\wll 6300, 6364$ doublet in SN 1987A during the first year indicated a volume filling factor of $\sim 0.1$ in the oxygen core \cite{Spyromilio1991,Li1992}, implying a factor ten higher density compared to the constant density  estimate above. For the iron-dominated region there is evidence of expansion from the radioactive heating during the first months after explosion \cite{Gabler2021}.  Modeling of the [Fe~{\sc ii}] optical and infrared (IR) lines indicated filling factors in the range $0.2 - 0.5$, despite constituting only $2.4 \%$ of the core mass  \cite{Li1993,Kozma1998}. The low density of the Fe dominated regions in combination with the strong emission lines from the radioactively powered ejecta could be responsible for the weak narrow [Fe~{\sc ii}] lines. As mentioned above, the
3D IFU observations 
of the [Si {\sc i}]/[Fe~{\sc ii}] $\wl 1.64$~{\textmu}m \ line \cite{Kjaer2010,Larsson2013}, the [Fe {\sc i}] $\wl \ 1.44$~{\textmu}m \ line \cite{Larsson2023},  and lines of CO and SiO  \cite{Matsuura2017},
indicate a highly asymmetric and clumpy distribution. This agrees qualitatively with hydrodynamical simulations of the explosion \cite{Gabler2021}, where the combination of an early reverse shock and ${}^{56}$Ni heating produces a clumpy structure with high-density regions mixed with low-density cavities. 
If a PWN is present, this would produce further clumping. Both analytical calculations \cite{CF1992} and hydrodynamical simulations \cite{Blondin2017} predict that the shell swept up by the pulsar wind is unstable and can lead to filamentation. The strongest effects are in the break-out phase when the bubble shell accelerates down the steep density gradient. Based on the observed filling factor above, $\sim 0.1$ , for the oxygen core above, we adopt a density ten times the uniform core density in Fig. \ref{fig:photcalc}. Models with higher filling factors are discussed in Sect. \ref{sec:core_density}.
 
There is also  evidence for mixing in velocity: H is mixed in from the envelope to the core \cite{Kozma1998,Larsson2019}, while Fe is mixed in the opposite direction  \cite{Spyromilio1990,Haas1990}. 
An example of a clumpy, anisotropic and strongly velocity mixed ejecta is Cas A, where knots and filaments formed in the explosion from different nucleosynthesis zones are completely mixed in velocity  \cite{Hughes2000,Milisavljevic2013}. 

The inhomogeneous, anisotropic  structure  close to the region where the central emission originates, and also the presence of dust, means that only a fraction of the ionizing luminosity is converted into optical and IR emission lines. The rest  either escapes the center and is absorbed at larger radii in the ejecta, or is converted to thermal radiation by dust. 
To model this we introduce a parameter representing the covering factor, $f$, of the line emitting regions as seen from the ionizing source, $f=\Omega/4 \pi$, where $\Omega$ is the solid angle. The predicted line luminosities are a factor $f$  lower compared to the spherical case. $f$ is a free parameter that we fit to the data so that the [Ar~{\sc ii}] $6.985$~{\textmu}m line agrees with the observations. This determines the luminosities of the other lines in the model from the relative line fluxes.  

\subsubsection{Shock models}
\label{sec:shocks}
We calculated models of line emission from shocks, with a setup corresponding to a PWN shock  (Sect. \ref{sec:origin}). The code is a major update of \cite{CF1994,Nymark2006}, similar to \cite{Williams2008}. We  solve the steady state hydrodynamic equations, including cooling and radiative emission in the same way as for the photoionization models, but including the time dependent terms in the ionization and thermal balance.
The atomic data is the same as used for the photoionization models. We do not include any ionization precursor in the model, as in \cite{Williams2008}, but have tested the sensitivity to the assumed state (here assumed to be in the third ionization stage, e.g. [O~{\sc iii}] etc.), finding that different assumptions have limited effects on the resulting line luminosities.  An inherent limitation to these one-dimensional models is that they are subject to instabilities \cite{Strickland1995,Sutherland2003}.  If the real geometry is not spherical or planar, a range of shock velocities would be produced by oblique shocks. Therefore we regard these models as indicative only. The code is written in Fortran and available on Zenodo \cite{ShockCode}, together with the output files.

\subsection{Relation to the dust peak}
\label{Sec:ALMA_peak}

Dust could thermalize emission from a compact object if the optical depth is high at optical-IR wavelengths \cite{Alp2018,Cigan2019,Page2020}. A submillimetre map of the dust temperature in SN 1987A \cite{Cigan2019} shows an increase in the temperature at a position slightly north of the center.  The position of the [Ar~{\sc vi}] centroid is located at $38\pm22$ mas east and $31\pm 22$ mas south of the center (section \ref{sec:spatial_ar}).  
The location of the dust peak was only quoted as the position of the brightest pixel, without an uncertainty \cite{Cigan2019} (though the offset of this pixel of 72 mas east and 44 mas north of the center was stated to be 3-5 times the total alignment uncertainty). This prevents us from  assessing the significance of the offset between the dust temperature peak and Ar emission site. If the dust peak is due to heating  by a CNS,  the CNS must be very close to the location of the peak \cite{Page2020}. The offset between that position and the Ar emission we observe is difficult to explain in this scenario.

The centroid of the [Ca~{\sc iv}] line at $3\pm10$~mas east and $114\pm15$ mas north from the center (section \ref{sec:lines_SM}) is closer to the dust peak than the [Ar~{\sc vi}] centroid, but any possible association between [Ca~{\sc iv}] and the dust emission cannot be quantified due to the lack of uncertainty on the dust peak.

\section{Supplementary text}
\subsection{Ejecta lines and comparison to other young supernova remnants}
\label{sec:comparison}
We only detect these the narrow emission line components from S, Ar and possibly Ca. 
The chemical composition in the center of the ejecta is likely dominated by elements resulting from  oxygen and silicon burning.
At the low temperature  of the high metallicity inner ejecta ($\lesssim 2000$ K, see below), ions which have low excitation temperature near infrared (NIR) and MIR transitions are predicted to dominate the cooling and therefore the observed spectrum. 

Both theoretical models and observations of other SNRs support this interpretation (see below).
The models predict that Si is the most abundant element in this region; the strongest of the lines from this element are the [Si ~{\sc i}]  1.607, 1.645~{\textmu}m and [Si~{\sc ii}] 34.81~{\textmu}m. However, Si ~{\sc i} is easily ionized, while the [Si~{\sc ii}] line is outside our observed wavelength range. Although the [S~{\sc iii}] 18.71~{\textmu}m and [S~{\sc iv}] 10.51~{\textmu}m  lines dominate the cooling of the Si -- S -- Ar -- Ca zone, their wavelengths are in a region of high dust absorption (main text) and are therefore expected to be weak. 

Several other young supernova remnants (SNRs) have been observed to show strong high-ionization lines with abundances characteristic of the inner ejecta, powered by synchrotron radiation from a PWN or thermal emission from shocks. 
The most extensively studied case is SNR Cas A, where  fast-moving knots are dominated by [S~{\sc ii}], [Ar~{\sc iii}] and [Ca~{\sc ii}], mixed with O lines with varying strengths, indicating ejecta that has undergone different
degrees of O burning \cite{Chevalier1979}. 
MIR Spitzer observations of SNR Cas A  have shown strong lines of [Ni~{\sc ii}] $6.636$, [Ar~{\sc ii}] $6.985$, [Ar~{\sc iii}] $8.991$, [S~{\sc iv}] $10.51$, [Ne~{\sc ii}] $12.81$, [Ne~{\sc v}] $14.32$, [Ne~{\sc iii}] $15.56$, [Fe~{\sc ii}] $17.94$, [S~{\sc iii}] $18.71$, [O~{\sc iv}] $25.89$, [Fe~{\sc ii}] $25.99$,  [S~{\sc iii}] $33.48$, and  [Si~{\sc ii}] $34.82$~{\textmu}m \cite{Smith2009}. Because SNR Cas A does not contain a PWN, these lines must be excited by either shocks or the ionizing radiation produced by the reverse shock. 

Both SNR G54.1+0.3 and SNR 0540-69.3 contain a PWN, with strong synchrotron emission. 
In the case of SNR G54.1+0.3 MIR spectra show ejecta lines of  H$_2$ \ 12.3, 17.0~{\textmu}m  and atomic lines from [Ar~{\sc ii}] $6.985$, [S~{\sc iv}] $10.51$, [Ne~{\sc ii}] $12.81$, [Cl~{\sc ii}] $14.37$, [S~{\sc iii}] $18.71$, [Fe~{\sc ii}] $25.99$ , [S~{\sc iii}] $33.48 $, and [Si~{\sc ii}] $34.82$~{\textmu}m \cite{Temim2010}.
The MIR $10-37$~{\textmu}m spectrum of SNR 0540-69.3 shows similar high-ionization lines of [O~{\sc iv}], [Ne~{\sc ii-iii}], [Si~{\sc ii}], [S~{\sc iii-iv}] and [Fe~{\sc ii}] \cite{Williams2008}.  

In all three cases the same set of atomic lines are seen in these remnants, with few variations. All three have a strong [Ar~{\sc ii}] line and strong lines above $\sim 8$~{\textmu}m from highly ionized heavy elements. While also SN 1987A has a strong [Ar~{\sc ii}] line, it differs from these SNRs by having weak or abscent 
lines above $\sim 8$~{\textmu}m. We discuss a possible explanation for this in the main text and in Sect. \ref{sec:dust}. 

\subsection{Origins of the narrow ejecta lines}
\label{sec:origin}
 
The observed displacement, $\sim -250 \kms$, of the [Ar~{\sc ii}] and [Ar~{\sc vi}] lines from the systemic velocity of SN 1987A, is much smaller than the general ejecta velocity, which reaches $\sim 5000 \kms$ close to the ER and $\sim 10,000 \kms$ at the reverse shock out of the ER plane \cite{Larsson2023}. Together with the spatial distribution, this indicates  that the Ar emission must have an origin close to the center. 
If the emission was coming from a spherically symmetric shell, this would have resulted in a boxy, flat-topped line profile, with a minimum and maximum velocity equal to the velocity of the shell, inconsistent with the observations. Instead we observe a displaced line with a FWHM width  $\sim 122 \kms$ of the [Ar~{\sc ii}] line.

There are several possible scenarios for the excitation of the narrow  ejecta
lines, including emission from a PWN  produced by a NS in the center, thermal emission directly from a CNS, accretion onto a compact object, shock excitation in the inner ejecta, excitation by  radioactive ${}^{44}$Ti,  a surviving companion star, dust reflection of the narrow line emission from the ER collision or emission from a reverse shock. We investigate each of these possibilities below.

Dust reflection from the ER is possible because [Ar ~{\sc ii}] is seen both in the ER and central ejecta.
Many silicates 
scatter light mainly  below $\sim 8$~{\textmu}m, but are strongly absorbing above this wavelength (Sect. \ref{sec:dust}).
However, if this was the origin of the [Ar ~{\sc ii}] lines we would expect similar reflections for the other strong lines from the ER below $\sim 8$~{\textmu}m, which we do not observe. 
Fig. \ref{fig:ar6_ringcomp} shows that a nearby H ~{\sc i} line at $\sim 4.376$~{\textmu}m is stronger than the [Ar~{\sc vi}] line in the ER spectrum but does not show a central emission.   Equivalent situations occur for many other strong lines in the NIRSpec range. In addition,  the spatial FWHM of the central component [Ar~{\sc vi}] line is $\lesssim 50\%$ of that of the ER and is blueshifted by over 200~km~s$^{-1}$ from the ER emission. 

More than $\sim 1500$ days after explosion, radioactive ${}^{44}$Ti  dominates energy input to the inner ejecta \cite{FK2002}. Modeling of the optical spectra \cite{Jerkstrand11} and observations of hard X-rays \cite{Boggs2015,Alp2021} have shown that $\sim 1.5 \times 10^{-4} \msun$ of ${}^{44}$Ti was produced in the explosion, which undergoes radioactive decay emitting gamma-rays and positrons. Most of the gamma-rays escape, while positrons are mostly trapped in the ejecta, where their annihilation provides an energy input of $\sim 1.2 \times 10^{36} \ergs$ \cite{Jerkstrand11}. If a large fraction of this energy was deposited in the Ar region of the ejecta, it could power the observed emission lines. However, the narrow line width indicates the emission must arise close to the center (see above). This is unlike the broad lines from the ejecta, such as the strong [Fe~{\sc ii}] $\wl 1.644$~{\textmu}m and $\wl 25.99$~{\textmu}m lines, which have a velocity of at least 2500 $\kms$ \cite{Larsson2023,Jones_overview_2023}.
Because of the low energy of the secondary electrons produced by the positron thermalization \cite{Kozma1992}, this process is unlikely to produce ions higher than singly ionized ions from the non-thermal positrons \cite{Jerkstrand11,Chugai1997}, so would produce very little [Ar~{\sc vi}] emission. While the broad components seen in the [Fe~{\sc i}] and [Fe~{\sc ii}]  lines arise from ${}^{44}$Ti and X-ray input from the ER collision \cite{Fransson2013}, we therefore conclude the ${}^{44}$Ti mechanism is  unlikely to be the source of the narrow Ar components. This was also the conclusion from a detailed study of the heating of the submillimter dust 'blob' \cite{Page2020}.

Another possibility is that the X-ray input from the ejecta and circumstellar  medium interaction could be responsible for the high ionization lines from the center. Models predict that a small fraction of the hard X-rays penetrate all the way to the central regions \cite{Fransson2013}. However, most of the luminosity in the X-rays from the shocks is emitted at energies below $\sim 1$ keV, so absorbed in the outer parts of the ejecta. The effect of the X-ray input can be seen in the [Fe~{\sc ii}] 5.340~{\textmu}m line, which is strong  especially the south-west region [\cite{Jones2023}, figures 15 and 16]. However, there is little emission in this line from the central region, indicating that X-rays from the ER do not penetrate to the centre of the ejecta. It is therefore unlikely that external X-ray flux could excite the narrow high-ionisation emission lines we observe in the center, although it could be responsible for some [Ar~{\sc ii}] emission from the region close to the ER.

A reverse shock is known to be present in SN 1987A, but only close to the ER, with a velocity of $\gtrsim 4000 \kms$ \cite{Larsson2023}. The inner ejecta must have been traversed by a reverse shock, but only within a few days of the explosion. Simulations have predicted the formation of an additional reflected shock, but it is expected to dissipate within a year \cite{Gabler2021}. We therefore exclude the possibility of reverse shock excitation.   

A surviving companion star is constrained to $\lesssim 3.7 \msun$ if on the main sequence \cite{Alp2018}, which corresponds to an A to F star. These produce few ionising photons and have soft spectra, insufficient to explain the [S~{\sc iii}], [S~{\sc iv}] and [Ar~{\sc vi}]  
emission. 

\subsection{Scenarios with a compact object}
\label{sec:compact_scenarios} 
Assuming no pulsar activity, the minimum bolometric luminosity of a central NS is that of a young CNS,  which has a temperature of $\gtrsim 10^6$ K and luminosity $\gtrsim 10^{34} \ergs$ \cite{Beznogov2021}, sufficient to ionize a substantial region of the inner core. 
We discussed this possibility in the main text and give more details below.

For the scenario with a PWN, discussed in the main text, we assume the PWN produces X-ray emission with a  power law spectrum with $L(\nu) \propto \nu^{-\alpha}$ with power law index $\alpha=1.1$, similar to the Crab PWN, normalized to a total luminosity above 13.6 eV less than the inferred upper limit of $10^{36} \ergs$ \cite{Alp2018}. 
From broad band X-ray observations of several PWN spectra there is  a  significant break in the power law at energies $4-14$ keV \cite{Madsen2015,Bamba2022}, consistent with a cooling break \cite{Chevalier2000}. However this has little effect on the line emission (Sect. \ref{sec:additional_models}). 

If a PWN is present, it produces synchrotron emission and a slow, dynamic shock that propagates into the ejecta  \cite{CF1992}. This shock is likely to be cooling, emitting line emission in the IR to soft X-ray range, and can therefore contribute to the emission from the inner ejecta. Because the pressure is approximately constant behind the shock, the cooling produce a high density, thin shell between the ejecta and PWN. The expansion velocity of this shell, $V_{\rm shell}$,
depends on the energy input by electromagnetic radiation, relativistic particles and magnetic fields from the pulsar to the PWN. We expect that to be close to the spin-down luminosity of the pulsar, $L_{\rm sd}$. 

To estimate $L_{\rm sd}$ we note that the ratio of the synchrotron luminosity to the spin-down luminosity is $\sim 28 \%$ for the Crab PWN \cite{Hester2008}. An upper limit to the bolometric luminosity from a compact object in SN 1987A is $\sim 10^{36} \ergs$ \cite{Alp2018}. Using this, we estimate a total spin down luminosity of $L_{\rm sd} \lesssim 3\times 10^{36} \ergs$. 
For an oxygen core mass $M_{\rm core} \approx 3 \msun$   and a core velocity $V_{\rm core} \approx 2200 \kms$ the kinetic energy 
of the core is $\sim 7.2 \times 10^{49}$ ergs. 
For a constant density core we then estimate $V_{\rm shell}$ [\cite{CF1992}, their equation 2.10] 
\begin{equation}
V_{\rm shell}= 275 \left(\frac{L_{\rm sd}}{3 \times 10^{36}  \ergs}\right)^{0.2}  \left(\frac{V_{\rm core}}{2200 \ \kms}\right)^{0.3} \left(\frac{M_{\rm core}}{3  \msun}\right)^{-0.2}  \left(\frac{t}{35  \ \rm years}\right)^{0.2}   \kms
\end{equation}
where $t$ is the time after the SN explosion. Alternatively, for a core with a density decreasing with radius, $r$ as $\rho \propto r^{-1}$: 
\begin{equation} 
V_{\rm shell}= 148 \left(\frac{L_{\rm sd}}{3 \times 10^{36}   \ergs}\right)^{0.25}  \left(\frac{V_{\rm core}}{2500 \ \kms}\right)^{0.5} \left(\frac{M_{\rm core}}{3  \msun}\right)^{-0.25}  \left(\frac{t}{35 \rm \ years}\right)^{0.25}   \kms.
\end{equation}
The PWN shock velocity into the expanding ejecta is in the two cases $V_{\rm shell}/6$ and $V_{\rm shell}/5$, respectively. The resulting shock velocities are then $46 \kms$ and $30 \kms$, respectively. The low shock velocity is because of the relative velocity of the shell and expanding ejecta. The uncertainties in the parameters are large, in particular these estimates assume a spherically-symmetric uniform core, which is a unlikely to be the case \cite{Blondin2017}. Nevertheless, we conclude that a PWN shell can produce line emission at low velocity.  

MIR observations of SNR G54.1+0.3 show  fine-structure emission  lines from [Ar~{\sc ii}] and other lines similar to other SNRs \cite{Temim2010,Temim2017}. This emission was interpreted as coming from  a slow shock, $\sim 25 \kms$, from the expanding PWN into the ejecta, producing a density enhancement. A similar model could in principle also explain our observations.  An expanding anisotropic PWN, as observed in  SNR G54.1+0.3, referred to as a  'jet' \cite{Temim2010}, could explain the fact that  we only observe narrow blueshifted emission lines. This can, however, also be due to internal ejecta dust absorption (see below). 

\subsection{Dust absorption and scattering}
\label{sec:dust}
Observations of dust in young, oxygen-rich SNRs, including Cas A \cite{Arendt2014} and SNR G54.1+0.3 \cite{Temim2017,Rho2018}  have been interpreted as showing different dust properties.
Modelling of the dust SED used mainly Mg$_{0.7}$SiO$_{2.7}$ and a secondary component of other silicates, carbon or alumina dust \cite{Temim2017}. Alternatively, it has been proposed that the $20$~{\textmu}m peak in SNR G54.1+0.3 is due to SiO$_2$ grains, while the $10$~{\textmu}m peak is produced by SiC grains and polycyclic aromatic hydrocarbons (PAHs) \cite{Rho2018}. The latter composition also matches the dust emission in Cas A \cite{Rho2018}. 

Other observations of Cas A have indicated different dust properties in the different abundance regions in the ejecta \cite{Arendt2014}. 
The Ar~{\sc ii} and Ar~{\sc iii}  
regions in the ejecta have strong silicate peaks, which were attributed to Mg$_{0.7}$SiO$_{2.7}$ \cite{Arendt2014}. 

Although there is  degeneracy in the models and conflicting interpretations of the dust spectra,
a common result is that silicates are invoked to explain  the $10-20$~{\textmu}m region. 
Because  dust formation occurs at high temperatures in the cooling ejecta,  a large fraction of the silicates can be in crystalline dust, as is seen in e.g., the outflows of asymptotic giant branch (AGB) stars and in comets \cite{Molster2005,Henning2010}. Whether the silicates are crystalline or amorphous depends on the cooling time scale compared to the expansion time scale. Fast cooling produces amorphous dust while very slow cooling favours crystalline.   

Independent of the exact dust composition,  size and structure of the grains, a common feature of silicates is a large drop of the absorption shortwards of  $8-9$~{\textmu}m.  Depending on the composition and crystallinity of the dust
this can be more or less sharp. We invoke this property to explain
the different MIR spectra of SN 1987A compared to other young SNRs, discussed in Sect. \ref{sec:comparison}, as well as the models discussed in this section.

\begin{figure}
    \centering
\includegraphics[width=13cm]{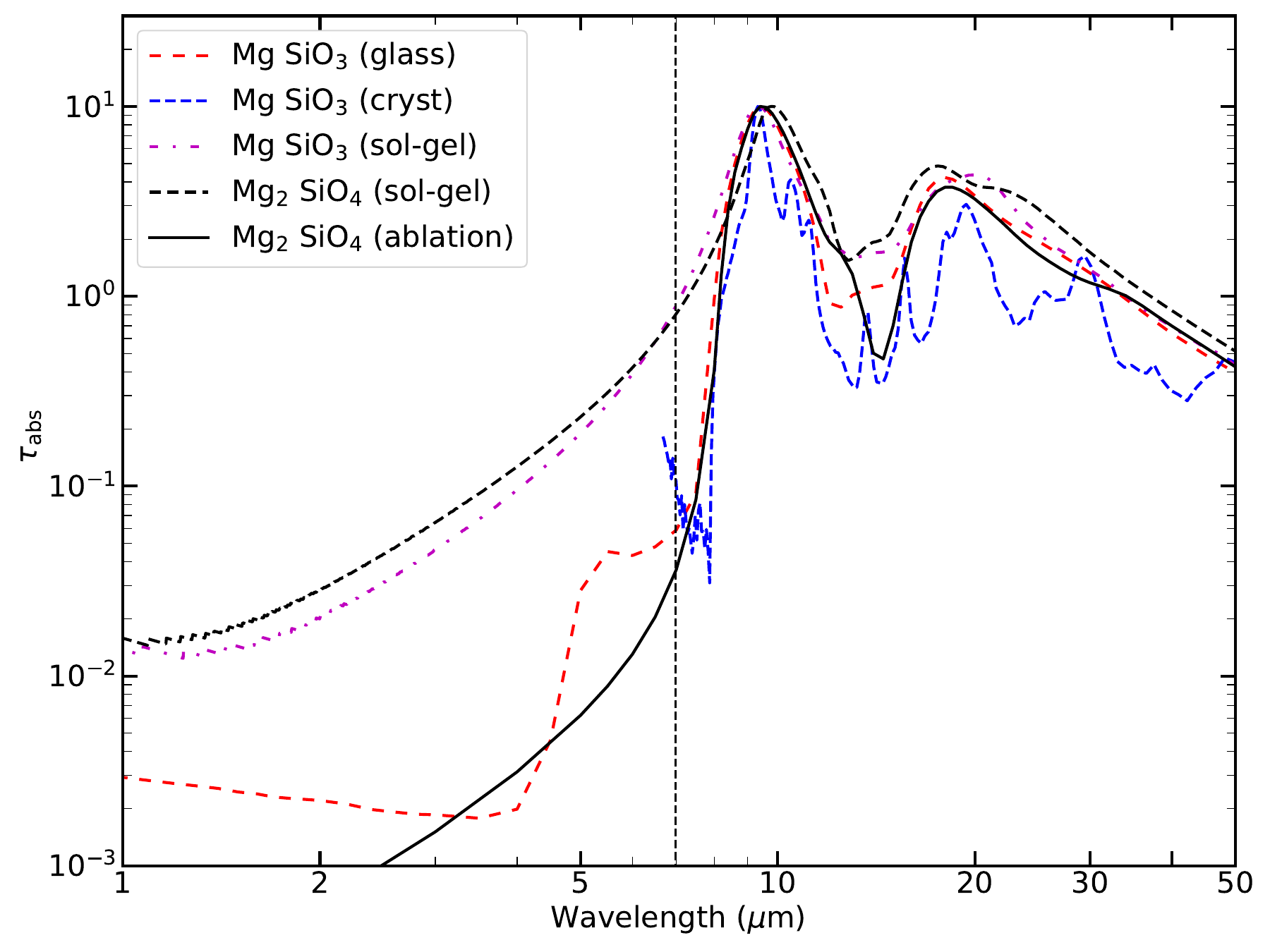}
\caption{\textbf{Optical depth of absorption as function of wavelength for different silicate compositions and structures.} The optical depths have been normalized to $\tau_{\rm abs} = 10$ at the $10$~{\textmu}m peak. There is a drop in $\tau_{\rm abs}$ below $\sim 8$~{\textmu}m in all cases. The wavelength of the [Ar~{\sc ii}] $6.985$~{\textmu}m line is shown by the vertical, dashed line. 
Optical constants of amorphous Mg$_2$SiO$_4$  obtained by laser ablation are from \cite{Tamanai2017,Gail2020} (solid line) and by the sol-gel method from \cite{Jager2003} (black dashed line). For enstatite, MgSiO$_3$, data for glass are from \cite{Dorschner1995}, for crystalline from \cite{Jaeger1994J} (blue dashed line) and  by the sol-gel method from \cite{Jager2003} (magenta dash-dotted line).}
\label{fig:sil_labs}
\end{figure}

Fig. \ref{fig:sil_labs} shows the  optical depth as a function of wavelength for several silicates, both in amorphous and crystalline forms, produced in laboratory experiments by different methods and therefore having different structures. The absorption is calculated in the Rayleigh limit, assuming grains are small compared to the wavelength and with the continuous  ellipsoidal distribution (CDE2) \cite{Min2003,Draine2021}. For comparison, we have normalized all optical depth curves to $\tau_{\rm abs}=10$ at $10$~{\textmu}m. Optical constants for amorphous Mg$_2$SiO$_4$ have been obtained using different techniques: laser ablation \cite{Tamanai2017,Gail2020} was used to produce the sample, or  the sol-gel method \cite{Jager2003}. Both methods lead to similar absorption curves above $\sim 15$~{\textmu}m, but  below the 10~{\textmu}m
peak the sol-gel method gives a flatter slope. Therefore  even for the amorphous case the optical properties depend on the structure for the same composition. A similar behaviour is seen for enstatite, MgSiO$_3$, where the MgSiO$_3$ data are from glass \cite{Dorschner1995} and sol-gel data \cite{Jager2003}. 
Observations of several AGB stars show a steeper slope below the 10~{\textmu}m
peak than amorphous silicates produced by the sol-gel method \cite{Jager2003}.
Crystalline enstatite  \cite{Jaeger1994J} has a similar steep slope below 10~{\textmu}m, but with considerable structure in the absorption above the peak. 

At wavelengths shorter than $\sim 8$~{\textmu}m dust scattering usually dominates absorption by silicates \cite{Dorschner1995}, especially at the shorter wavelengths. The scattering and absorption properties depend strongly on the size of the grains, as well as on composition, especially if Fe is included, e.g. MgFeSiO$_4$ 
\cite{Dorschner1995}. The grain size is also important, with larger grains increasing the scattering. Theoretical calculations indicate a  distribution that peaks at $\sim 0.1$~{\textmu}m, extending to $\sim 1$~{\textmu}m \cite{Sarangi2015}. 

Pure scattering does not thermalize the photons, but produces broader lines from the expanding ejecta and more spatially extended emission that is  difficult to separate from the general ejecta emission. 
If there is optically thick dust in the core, mixed with the line emitting clumps, emission from the far side  relative to the observer (redder velocities) could be  absorbed, as was first proposed for SN 1987A as an explanation of the blue shift of the emission lines during the first years \cite{Lucy1989}. This, together with clumping, could explain the  blueshift of the Ar lines we observe. This requires the dust to have some absorption below 8~{\textmu}m, which is the case for olivine, MgFeSiO$_4$ \cite{Dorschner1995}.

\subsection{Detailed results for the photoionization models in the main text}
\label{sec:model_results}
Figure \ref{fig:photcalc} shows the results of our 'standard' photoionization models for the PWN and CNS cases.
For these models   
with $n_{\rm ion}=2.6\times 10^4 \ccm$, we find that $\xi \approx 0.5$ gives the best fit to the line ratios for both models. Specifically, we used for the PWN model $L_{\rm ion}=5\times 10^{34} \ergs$ and $r=2.8 \times 10^{15}$ cm and $L_{\rm ion}=3\times 10^{35} \ergs$ and $r=5.6 \times 10^{15}$ cm for the CNS model. However, only the flux, $L_{\rm ion}/4 \pi r^2$, affects the results.

\subsubsection{Temperature and ionization structure}
\label{sec:temp_ionization}
\begin{figure}
\centering
\includegraphics[width=7.9cm]{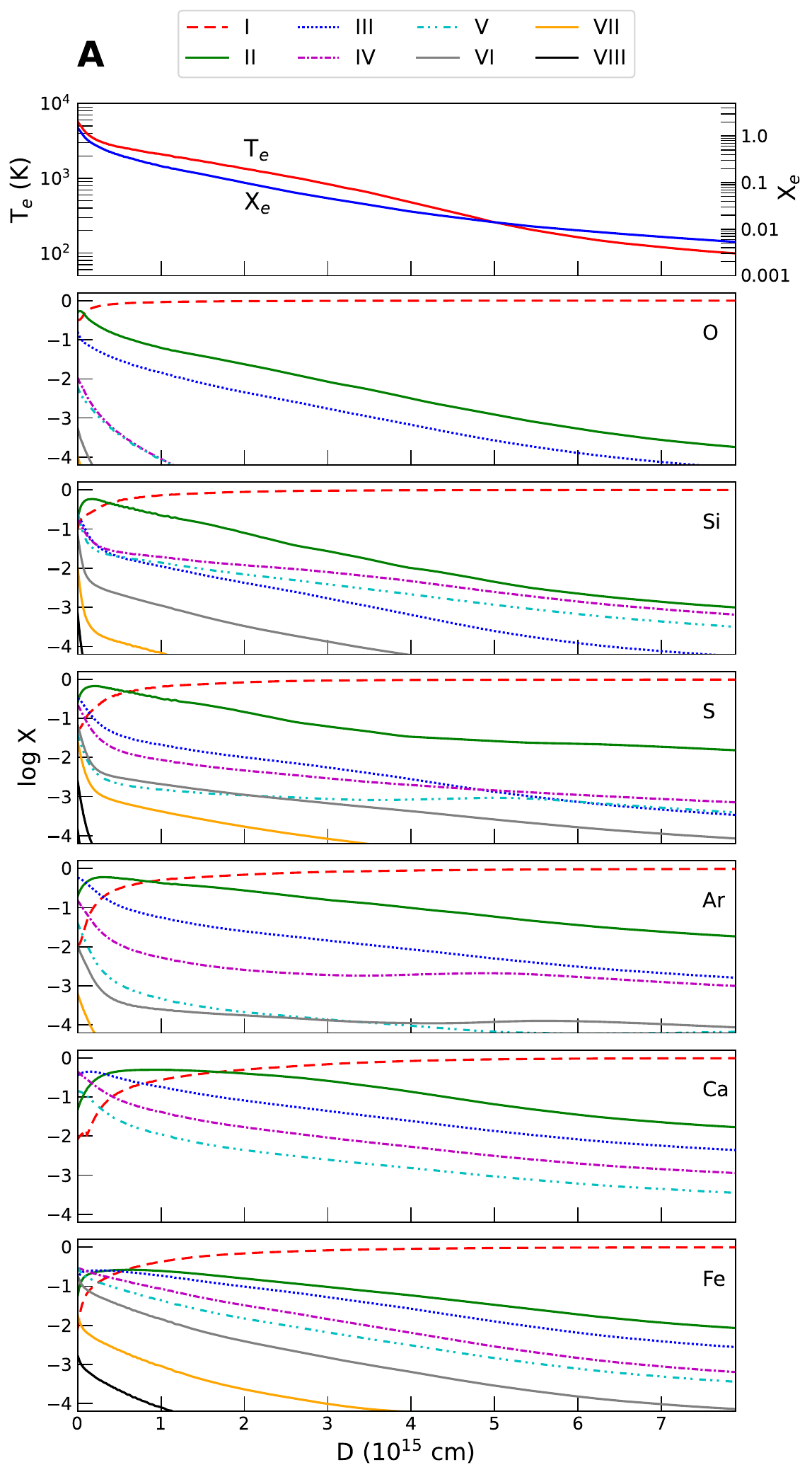}
\includegraphics[width=7.9cm]{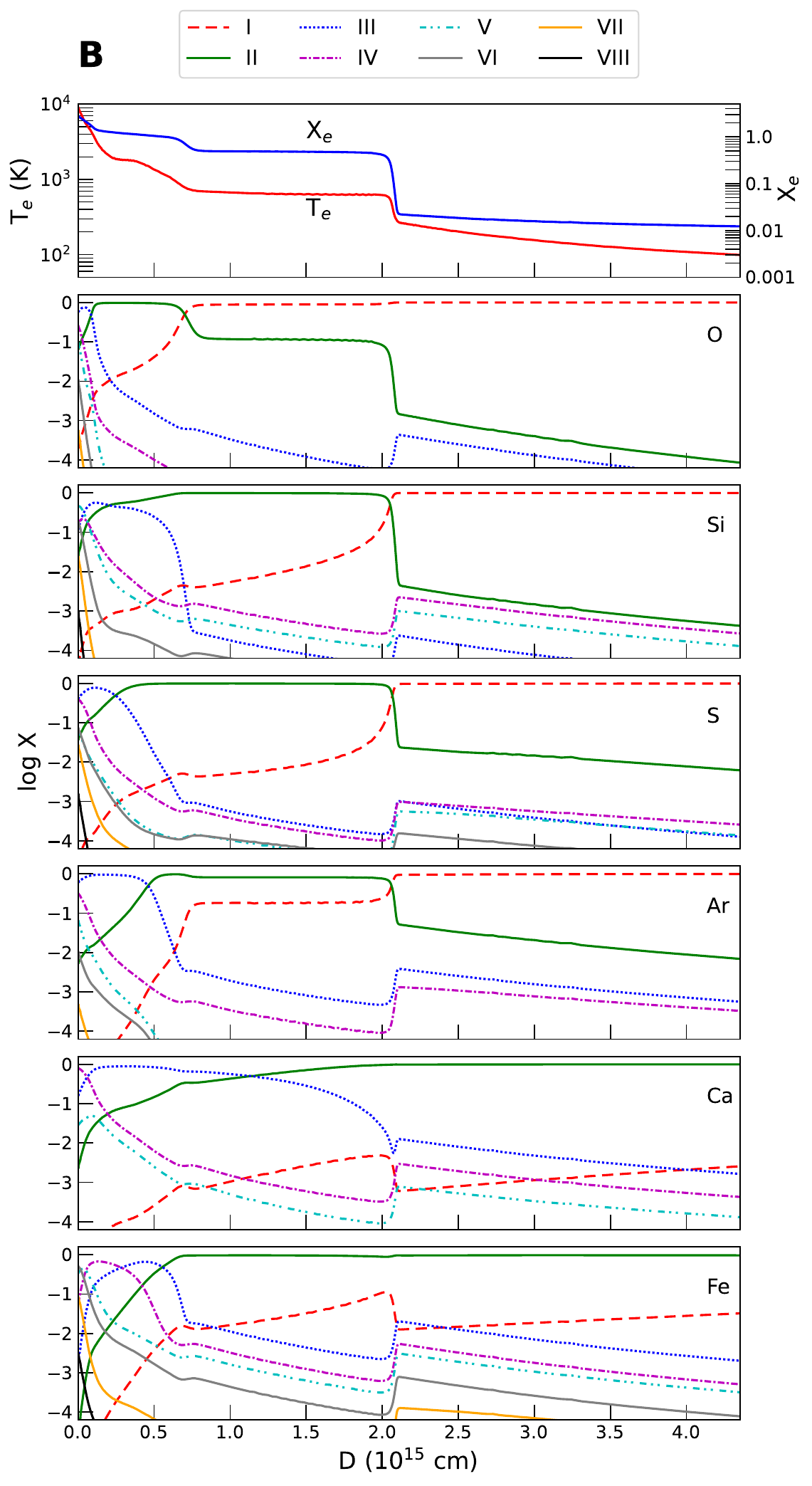}
\caption{\textbf{Photoionisation models of the
CNS (A) and a PWN (B) cases.} In the top panel the temperature, $T_{\rm e}$, is shown with the red curve and electron fraction, $X_{\rm e}$, with the blue. The different ionization fractions, $X$, are color and line coded as shown by the legend. There are distinct ionization zones for the PWN, which overlap in the CNS case. 
}
\label{fig:ion_struc}
\end{figure}

Figure \ref{fig:ion_struc} shows the temperature and ionization structure of selected elements for the two cases. 
The ionization structure of the CNS and PWN models are quite different as a result of the different ionizing spectra. The blackbody like CNS spectrum leads to diffuse zones with a large range of ionization stages at the same radius, while the steeper PWN spectrum produces  more discrete zones. The high metallicity leads to a general lower temperature compared to a gas with solar abundances, $\lesssim 10^4$ K in the high ionization zone, decreasing to a few hundred K in the partially ionized zone. A consequence of this is that most of the cooling is dominated by NIR and MIR fine structure lines of O~{\sc i-iii}, Si~{\sc i-ii}, S~{\sc i-iv}, Ar~{\sc ii-v}, with low excitation temperatures. 

The thickness of the ionized region is determined by the number of ionizing photons, electron density and recombination rates of the dominant ions, and the ionization parameter, so $\Delta r \propto \xi$.   
Figure \ref{fig:ion_struc} shows the thicknesses in both models are  $\lesssim 10^{16}$~cm, which is too small to resolve in our observations. 

Compared to  photoionization regions with solar abundances, the higher abundance of heavy elements, which are efficient coolants, results in a lower temperature, $\lesssim 1000$ K, in most of the line emitting region, even if the ionization is higher. The result is that  few of the usually strong UV resonance and semi-forbidden lines seen in objects with solar abundances can be excited, and the cooling and emission is instead dominated by MIR and far-IR lines from heavy elements. Only close to the inner boundary, 
and for a high ionization parameter, is there a zone with temperature $\gtrsim 10^4$ K where UV lines are excited. 

In addition to the lines discussed in the main text, the photoionisation models predict other strong lines at shorter and longer wavelengths. Longwards of the NIRSpec and MRS range the models predict strong fine structure lines, 
including [Si~{\sc ii}] 34.81, [S~{\sc iii}] 33.49, [O~{\sc iii}] 51.81, 88.33~{\textmu}m. However, no observations are available in this range. In the optical, the models predict strong  [O~{\sc iii}]   4959, 5007 \AA, [Ar~{\sc iv}] 4712, 4741 \AA, [Ar~{\sc v}] 6435, 7006  \AA,  [Ca~{\sc ii}] 7291, 7323 \AA,  [Ar~{\sc iii}] 7136, 7751 \AA \ , with line ratios that differ between the models. However, dust scattering could affect these predictions.

\subsubsection{Lines in the NIR}
\label{sec:NIR_lines}
In the NIRSpec range  [Si~{\sc i}] $1.099, 1.607, 1.645$~{\textmu}m lines, [S~{\sc i}] $1.033$~{\textmu}m and [S~{\sc iii}] $0.9533$~{\textmu}m lines are predicted. These lines are 
stronger in the CNS models, due to the extended Si~{\sc i} zone with a temperature $\gtrsim 10^3$ K (Fig. \ref{fig:ion_struc}), which is high enough for collisional excitation of these lines.  The PWN models have more well-defined ionization zones, and a much lower temperature in the Si~{\sc i} zone, resulting in weak [Si~{\sc i}] lines.
We do not find any narrow lines at these wavelengths in the NIRSpec spectrum, which may speak in favour of the PWN model. The limited spectral resolution of NIRSpec and overlapping broad ejecta lines, however, makes it difficult to distinguish these.
Dust scattering can also broaden the lines both in wavelength and spatially.  
The low ionization potential of Si~{\sc i}, 8.15 eV, means that any radiation below 1520 \AA \ can ionize Si~{\sc i}. This effect is seen in the PWN models, which have a strong continuum at these wavelengths. The strong UV radiation from the ER shocks, including Lyman $\alpha$ \cite{Kangas2022}, will have the same effect, unless dust extinction to the center absorbs the radiation from the ER. There is also a strong UV field at these energies from the ejecta itself in the form of Lyman $\alpha$ and two-photon emission from hydrogen and helium powered by ${}^{44}$Ti decay \cite{Jerkstrand11}.  

The main difference between the models and observations in the NIRSpec range are the strong [Ca~{\sc iv}] 3.207~{\textmu}m and [Ca~{\sc vi}] 4.159~{\textmu}m lines. These lines are in the models an order of magnitude stronger than the [Ar~{\sc vi}] line in the same wavelength range. The [Ca~{\sc iv}] 3.207~{\textmu}m is observed from the central region (Fig. \ref{fig:line_prof_CaIV_VI_SM}), but is coming from a different location and is redshifted. 
This discrepancy could be caused by uncertainties in atomic data, in particular photoionization cross sections, collision strengths and  recombination rates. The Ca~{\sc iv} and Ca~{\sc v} ionic fractions are at least an order of magnitude larger than the Ar~{\sc vi} abundance (Fig. \ref{fig:ion_struc}). These effects could lead to overestimates of the Ca line strengths.  The relative abundances between calcium and argon depend on the ZAMS mass \cite{WoosleyHeger2007} and vary within the O--Si--S--Ar--Ca zones (Fig. \ref{fig:abundances}), but only within a factor of about two. The relative Ar to Ca abundances in this zone for models between 15 to 19 $\msun$, vary by a similar amount \cite{WoosleyHeger2007}. Adopting a 15 $\msun$ model gives similar results as for the 19 $\msun$ model. However, there could be systematic uncertainties in the treatment of both the pre-SN burning and the explosive oxygen burning in these one-dimensional models, which could produce incorrect different abundance ratios \cite{Muller2020LRCA}.

An alternative explanation for the high Ca line luminosities in the model could be the neglect of depletion of Ca to dust. Ca is not included in the dust models in \cite{Sarangi2015} nor \cite{Sluder2018}, so no quantitative predictions are available. However, calciosilia, like wollastonite, CaSiO$_3$ and diopside, MgCaSi$_2$O$_6$ \cite{Rietmeijer2008}, could be produced by the dust formation in the inner ejecta, , lowering the Ca abundance in the gas phase. The same applies to the NIR  [Si~{\sc i}] lines, which are predicted to be strong in the CNS model but not observed. The large mass of silicates indicated by the dust observations, and formation of molecular SiO \cite{Matsuura2017,Cigan2019}, could lower the gas-phase abundance of atomic Si. 

\subsubsection{The covering factor}
\label{sec:covering}
The ionization structure, and resulting line ratios, are determined by the ionization parameter, $\xi=L_{\rm ion}/n_{\rm ion} r^2$, for the same spectral shape, $L(\nu)$. 
Because we do not know how large the solid angle of the emitting gas is from the ionizing source we normalize the [Ar~{\sc ii}] $6.985$~{\textmu}m  luminosity in the model to the observed value by a covering factor $f=L_{\rm obs} ({\rm [ Ar~\textsc{ ii}]})/L_{\rm model}({\rm [Ar~{\textsc {ii}}]})$.
In the models $\sim 20 \%$ of  $L_{\rm ion}$ emerges in the [Ar~{\sc ii}] line, so
$f \approx 5 \ L_{\rm obs}$([{\rm Ar~{\sc ii}}])$ / L_{\rm ion}$. The ionization parameter is determined by the relative line luminosities, primarily the [Ar~{\sc i
vi}]/[Ar~{\sc ii}] ratio. Therefore we can write $f \approx 5 \ L_{\rm obs}$([{\rm Ar~{\sc ii}}])$ / \xi n_{\rm ion} r^2$, with $\xi \approx 0.2-0.4$. This means that $f$ is determined by the distance between the ionizing source and emitting gas which is not known, except for an upper limit. 

The luminosity, distance to the ionizing source and density of the  PWN model in Fig. \ref{fig:photcalc} were chosen to be $L_{\rm ion} = 5 \times 10^{34} \ergs$, $r=2.8 \times 10^{15}$ cm and $n_{\rm ion}=2.6 \times 10^4 \ccm$ for the PWN model, resulting in a covering factor    $f=0.022$. 
To produce the same ionization parameter for the CNS model with the same density and a luminosity $L_{\rm ion} = 3 \times 10^{35} \ergs$, indicated by the CNS models \cite{Page2020}, the distance has to be $5.6 \times 10^{15}$ cm, which results in $f=0.005$. For a lower NS luminosity the scaling is the same as above, $f \propto L_{\rm ion}^{-1}$, but the luminosity is more constrained to $\gtrsim 3 \times 10^{34} \ergs$ \cite{Page2020} than for the PWN case.

These low covering factors indicate that only a fraction of the  total ionizing luminosity is converted into the observed lines, and/or a lower luminosity and distance. 
A small covering factor is consistent with other aspects of our observations.
The observed lines observed are shifted with respect to the rest velocity and also narrow in velocity, 
which shows that only a small solid angle is responsible for the observed luminosity.
The total luminosity of the line emission is less than $\sim 10^{33}$ erg s$^{-1}$, which is at least two orders of magnitude less than  expected for a CNS \cite{Beznogov2021,Beznogov2023}.  
The difference may  be due to line emission, similar to the observed, but absorbed by optically thick dust in the NIR. Alternatively, the continuum flux from the ionizing source could be directly thermalized by  dust or absorbed by bound-free absorption in the ejecta. This would result in broad, low ionization lines, which would not be distinguished from the radioactively and X-ray powered lines, like the [Si~{\sc i}] and [Fe~{\sc ii}] lines in the NIRSpec and MRS ranges \cite{Larsson2023,Jones_overview_2023}. In the estimate of the observed total line emission above we have only included the narrow component of the lines in Fig. \ref{fig:line_prof}, while the broader component also contributes to the total [Ar~{\sc ii}] luminosity. 

\subsection{Alternative PWN models}
\label{sec:parameters}
\subsubsection{Sensitivity to the core density}
\label{sec:core_density}
\begin{figure*}
\begin{center} 
\includegraphics[width=14.cm]{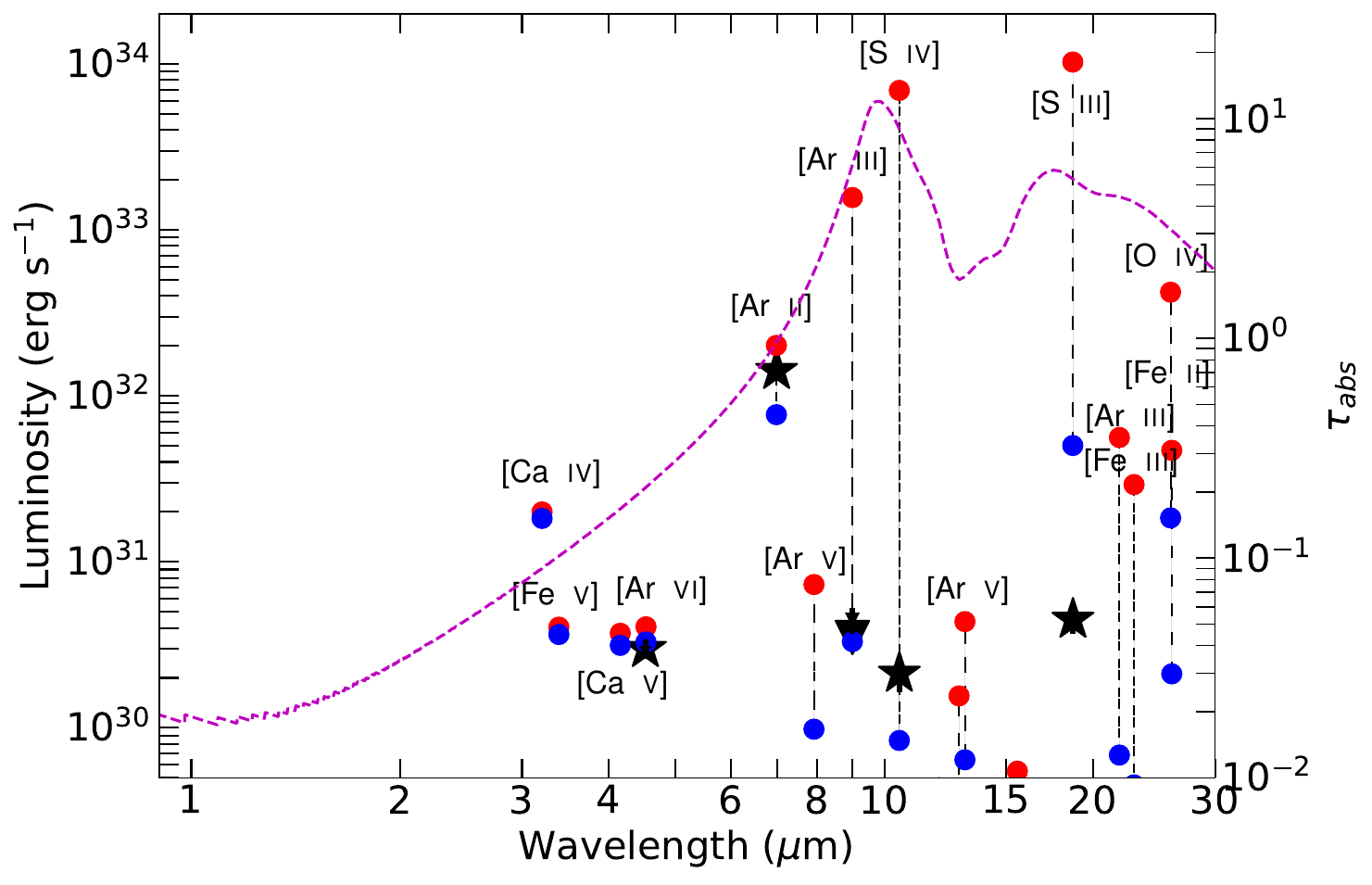}
\end{center}
\caption{\textbf{PWN photoionization model without clumping}. Same as Fig. \ref{fig:photcalc}B but assuming a ten times lower density, corresponding to a constant density core.  }
\label{fig:photcalc_avdens}
\end{figure*}

As discussed above, the density is uncertain because of clumping and filaments. Our models in Fig. \ref{fig:photcalc} assumed a factor of ten higher density than the average in the core. To check the sensitivity to this we calculated a PWN model with the average density, $n_{\rm ion}=2.6 \times 10^3 \ccm$. To match the observed [Ar~{\sc vi}] /[Ar~{\sc ii}] ratio, the ionization parameter, $\xi$, has to be within $\sim 50\%$ of that of the the models with the higher density. This requires  the ionizing luminosity to be reduced by the same factor as the density, or alternatively the distance increased by the square root of the density, or $1\times 10^{34} \ergs$.
The results of this model are shown in Fig. \ref{fig:photcalc_avdens}. 

Comparing Figures \ref{fig:photcalc} and \ref{fig:photcalc_avdens}, we find that the  [Ar~{\sc vi}] /[Ar~{\sc ii}] ratio is the same, as was expected. While qualitatively most of the MIR line luminosities relative to the [Ar~{\sc ii}] line are similar, quantitatively the [S~{\sc iii}] $  18.71$~{\textmu}m line is considerably stronger in the model, caused by the more extended S~{\sc iii} zone in the low density model. The fairly strong [S~{\sc iii}] lines at $0.90-0.95$~{\textmu}m in the high density model are absent in the low density model, which is a result of a lower temperature in the  S~{\sc iii} zone. Similarily, the weaker [S~{\sc iv}] $10.52$~{\textmu}m line is  a result of the lower temperature in the S~{\sc iv} zone.  

The main effect of the density in these models is that collisional de-excitation of the NIR and MIR fine-structure lines becomes important as the density approaches the critical density of these lines, $n_{\rm crit}=A_{\rm u,l}/C_{\rm u,l}$, where $A_{\rm u,l}$ is the radiative transition rate and $C_{\rm u,l}$ is the collisional de-excitation rate between upper level $u$ and lower level $l$. Typically $n_{\rm crit} \approx 10^4 \ccm$ for these lines. Collisional de-excitaion results in a lower cooling rate and therefore a higher temperature as the density increases towards $n_{\rm crit}$, explaining the weaker optical and NIR lines, with high excitation temperatures, in the high density models.  

\subsubsection{PWN models with different ionizing spectra }
\label{sec:additional_models}
X-ray observations have been interpreted \cite{Greco2022} as indicating a non-thermal steep power law with $L(\nu) \propto \nu^{-\alpha}$ with $\alpha=1.8$ in hard X-rays from a PWN [however the same observations have also been interpreted as indicating thermal emission \cite{Alp2021}].   A bolometric limit is 138 solar luminosities or $5.4\times 10^{35} \ergs$ for a compact object in SN 1987A \cite{Alp2018}. In \cite{Greco2022}  X-ray observations between 10-20 keV are fitted by a steep spectrum with $\alpha=1.8$ and 10-20 keV  luminosity $3.0\times 10^{34} \ergs$. Assuming that the $\alpha=1.8$ spectrum applies to 13.6 eV, the  luminosity between 13.6 eV to 20 keV is $1.4\times 10^{37} \ergs$, i.e., a factor of more than twenty higher than the limit above. Extending into the UV and optical would increase this discrepancy. To comply with the power law and luminosity from \cite{Greco2022} and the limit above, there must therefore be a break to a flatter spectrum at soft X-ray energies. 

We calculated two PWN models with this power law and luminosity between $10-20$ keV \cite{Greco2022}. The  density and ionization parameter are the same as the  PWN model in Fig. \ref{fig:photcalc}. To produce a luminosity of $3.0\times 10^{34} \ergs$ between 10-20 keV and a continuous spectrum requires $L_{\rm ion} \gtrsim 5 \times 10^{35} \ergs$ for $\alpha=1.1$ below the break. With $L_{\rm ion} = 1 \times 10^{36} \ergs$ the break to the $\alpha=1.8$ spectrum occurs at $E_{\rm b}=4.1$ keV, while $L_{\rm ion} = 2 \times 10^{36} \ergs$ gives $E_{\rm b}=1.1$ keV. 
\begin{table}[t!]
\centering
\caption{\textbf{Luminosities relative to the $[$Ar~{\sc ii}$]$ 6.985~{\textmu}m line for different power law breaks.} No correction for dust absorption has been applied.
}
\vspace{3mm}
\footnotesize{
\begin{tabular}{|l c l l l |} 
 \hline
Ion &\phantom{1}Wavelength\phantom{1234}&No break\phantom{111}&$E_{\rm b}=4.1$ keV&$E_{\rm b}=1.1$ keV\\    
&{\textmu}m&&&   \\
 \hline\hline
 &&&& \\ 
\vspace{-3mm}
$[$S~{\sc iii}$]$&    \phantom{1}0.9530&   0.9584&      0.8841&      0.9398\\
&&&&\\
\vspace{-3mm}
$[$C~{\sc i}$]$&    \phantom{1}0.9824&     0.2444&      0.2239&      0.2400\\
&&&&\\
\vspace{-3mm}
$[$Ca~{\sc iv}$]$&    \phantom{1}3.2068&     0.0966&      0.0930&      0.0995\\
&&&&\\
\vspace{-3mm}
$[$Ca~{\sc v}$]$&    \phantom{1}4.1585&     0.0136&      0.0133&      0.0139\\
&&&&\\
\vspace{-3mm}
$[$Ar~{\sc vi}$]$&    \phantom{1}4.5292&     0.0132&      0.0129&      0.0137\\
&&&&\\
\vspace{-3mm}
$[$Ar~{\sc ii}$]$&    \phantom{1}6.9850&     1.0000&     1.0000&     1.0000\\
&&&&\\
\vspace{-3mm}
$[$Ar~{\sc v}$]$&    \phantom{1}7.9141&     0.0213&      0.0207&      0.0221\\
&&&&\\
\vspace{-3mm}
$[$Ar~{\sc iii}$]$&   \phantom{1}8.9914&   0.9455&      0.9209&      0.9698\\
&&&&\\
\vspace{-3mm}
$[$S~{\sc iv}$]$&   10.5077&     0.7319&      0.7110&      0.7574\\
&&&&\\
\vspace{-3mm}
$[$Ar~{\sc v}$]$&     13.0680&     0.0081&      0.0079&      0.0084\\
&&&&\\
\vspace{-3mm}
$[$S~{\sc iii}$]$&   18.6992&     0.7328&      0.7160&      0.7533\\
&&&&\\
\vspace{-3mm}
$[$Ar~{\sc iii}$]$&   21.8315&     0.0346&      0.0340&      0.0354\\
&&&&\\
\vspace{-3mm}
$[$Fe~{\sc iii}$]$&   22.9192&     0.0057&      0.0056&      0.0061\\
&&&&\\
\vspace{-3mm}
$[$Ar~{\sc iv}$]$&   25.8800&     0.0299&      0.0292&      0.0314\\
&&&&\\
\vspace{-3mm}
$[$Fe~{\sc ii}$]$&   25.9814&     0.0344&      0.0324&      0.0320\\
&&&&\\
\vspace{-3mm}
$[$S~{\sc ii}$]$&   34.8100&      0.9410&      0.8834&      0.8965\\
&&&&\\
\hline
\end{tabular}
}
\label{table:2}
\end{table}
In Table \ref{table:2} we show the line luminosities relative to the $[$Ar~{\sc ii}$]$ 6.9850~{\textmu}m line for these two models, together with the model with no break (Fig. \ref{fig:photcalc}). The change in the luminosities for these lines are less than $10 \%$, and shows that a break above $\sim 1$ keV has only marginal significance for the IR emission line spectrum.  
This is because the ionization structure is mainly determined by the  spectrum below $\sim 1$ keV and the harder X-rays have little effect for $\xi \sim 0.5$.

\subsubsection{PWN models from the O -- Ne -- C-- Mg zone }
\label{sec:O_Ne_C_Mg_models}
\begin{figure*}[t!]
\begin{center} 
\includegraphics[width=16.cm]{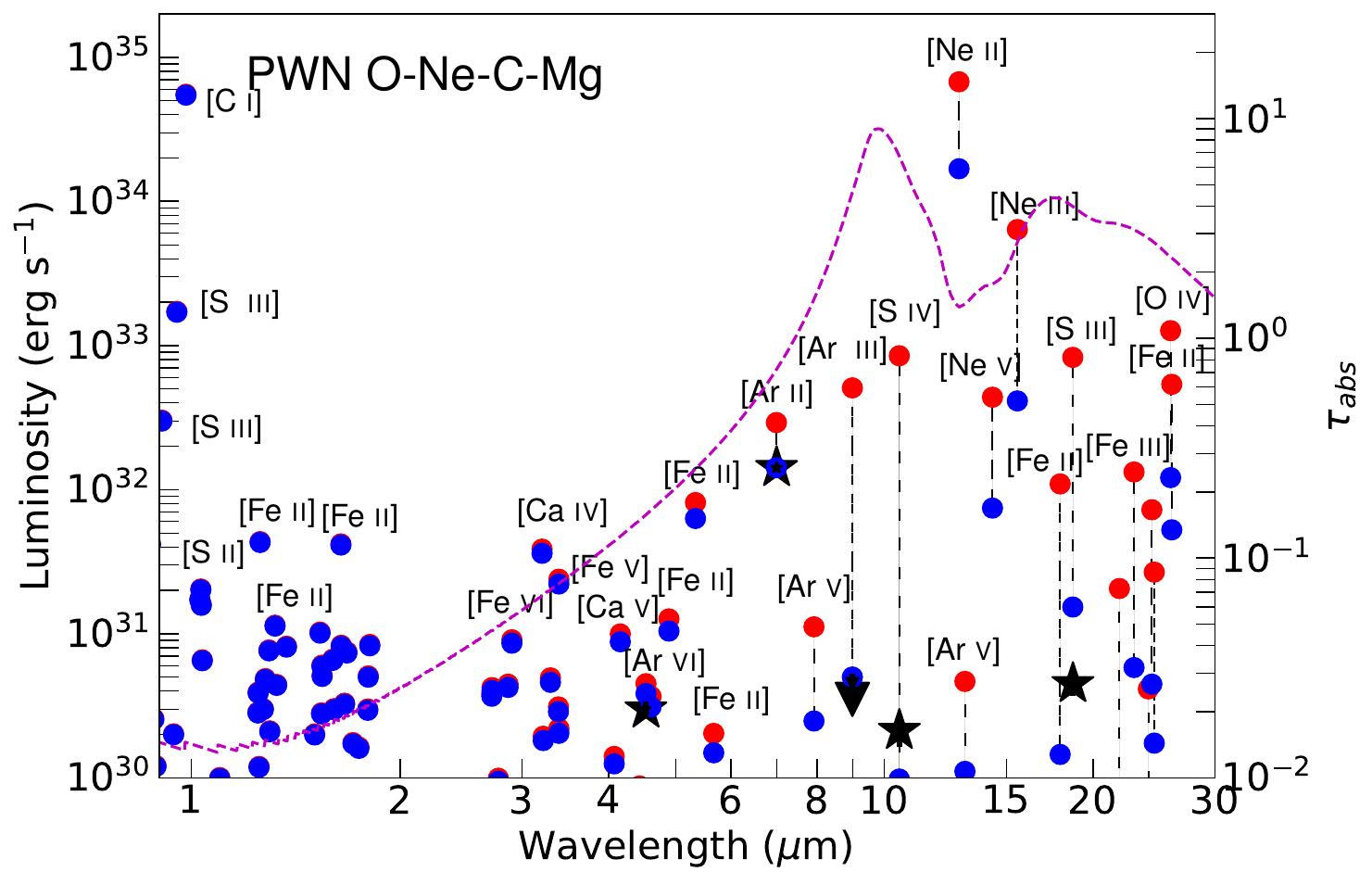}
\end{center}
\caption{\textbf{PWN photoionization model for the O-Ne-C-Mg zone.} Same symbols and colors as in Fig. \ref{fig:photcalc}}
\label{fig:photcalc_O_Ne}
\end{figure*}
We calculated models with abundances from the next nuclear burning zone, mainly O, Ne, C and Mg (Fig. \ref{fig:abundances}). The cooling is dominated by C~{\sc i}, O~{\sc i} - O~{\sc iii} and Ne~{\sc ii} - Ne~{\sc v}. The predicted spectrum (Fig. \ref{fig:photcalc_O_Ne}) is therefore very different compared to Fig. \ref{fig:photcalc}, with weaker lines of Ar, but instead strong lines of C, O, Ne, S due to the differing abundances (Table \ref{table:abund}). This is consistent with previous models \cite{CF1992}. Although the relative strengths of the [Ar~{\sc ii}] and [Ar~{\sc ii}] lines can be reproduced, this model can from the more than two order of magnitude stronger [Ne~{\sc ii}], [Ne~{\sc iii}] and [Ne~{\sc v}] lines be excluded as the origin of the observed narrow lines.

\subsection{Results for shock models}
\label{sec:shock_results}

\begin{figure}
\centering
\includegraphics[width=10.cm]{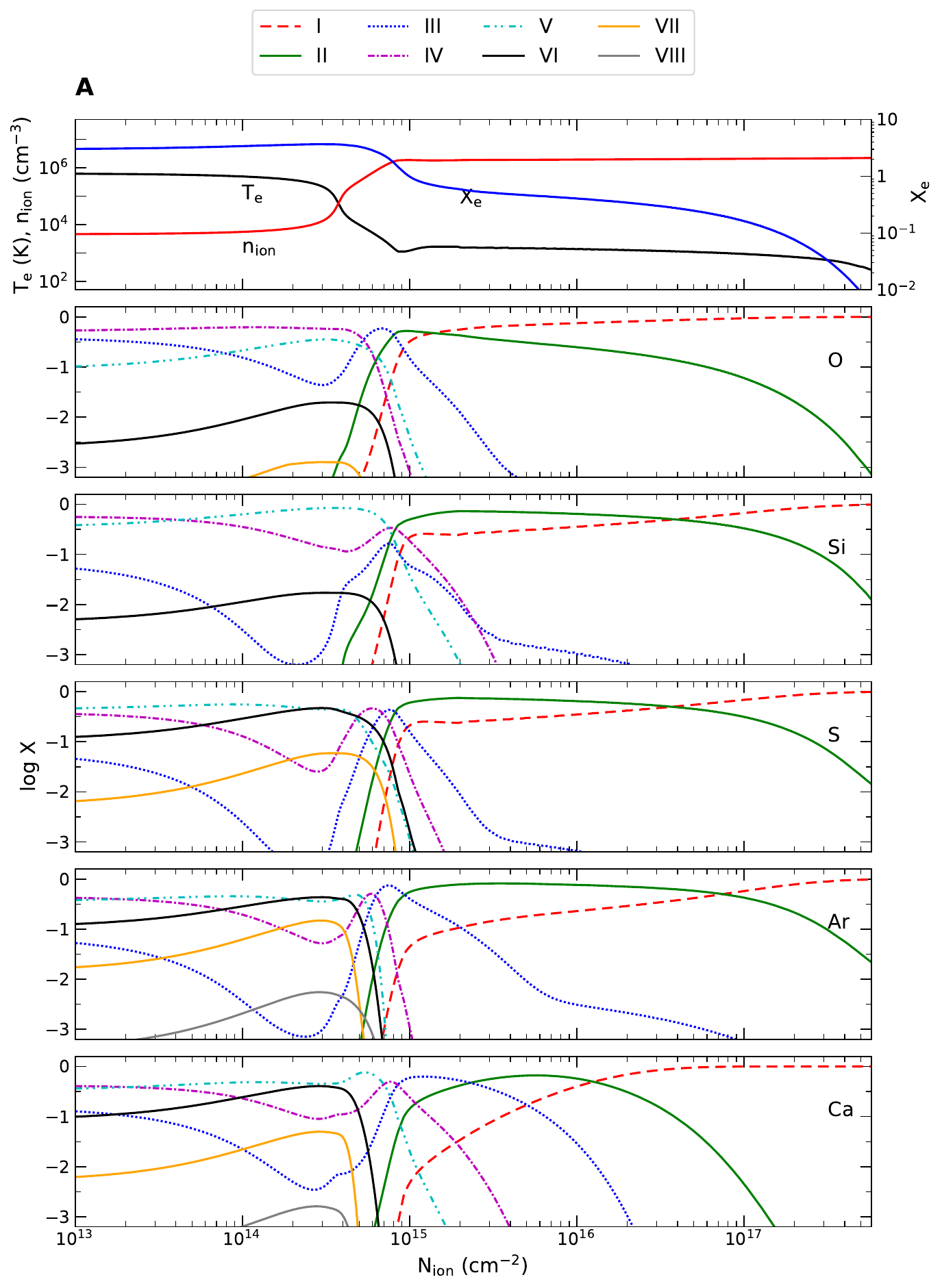}
\includegraphics[width=9.cm]{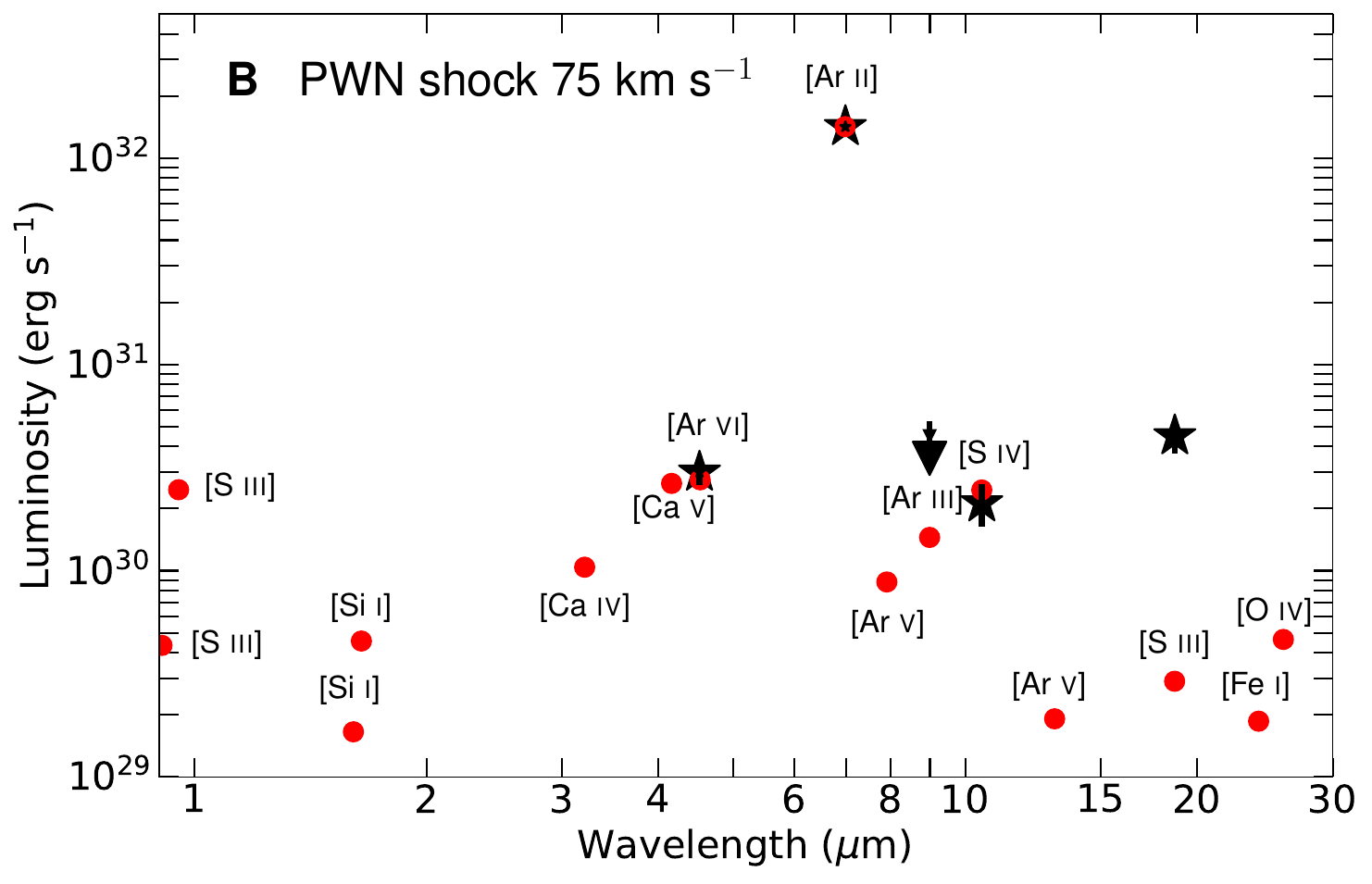}
\caption{\textbf{Structure (A) and predicted line luminosities (B) from a 75~km~s$^{-1}$ PWN shock}.  Line styles and colours indicate ionisation states (see legend). No dust absorption  has  been applied.
The symbols are the same as in Fig. \ref{fig:photcalc}.}
    \label{fig:pwn_shock}
\end{figure}
We calculated a model with parameters close to estimates for a PWN shock (Sect. \ref{sec:origin}),
$V_{\rm s}= 50\kms$  and pre-shock density $\rho_{\rm pre-shock} = 2 \times 10^3 \ccm$. This model, however, predicts  an  [Ar~{\sc vi}]/[Ar~{\sc ii}]  ratio more than an order of magnitude lower than observed, and also underpredicts the other high ionization lines. To get better agreement a higher velocity is required. 
As discussed in Sect \ref{sec:ab_struc}, the ejecta is expected to be clumpy. If the pre-shock density is lower than the average used above, the shock velocity will be higher, $V_{\rm s} \propto \rho^{-1/2}_{\rm pre-shock}$ for the same bubble pressure. 
Fig. \ref{fig:pwn_shock} shows the result from a model with density, $\sim  1 \times 10^3 \ccm$ and $V_{\rm s}= 75 \kms$, which more closely matches the observations, especially the high ionization lines. This model does not include any ejecta absorption, because it would completely extinguish the already  low luminosities of lines above $\sim 8 \ \mu$m relative to the [Ar~{\sc ii}] line. 

Fig. \ref{fig:pwn_shock}A shows the resulting temperature, density and ionization for this model 
as a function of the column density of the ions, $N_{\rm ion}$.
The model predicts an almost constant temperature, $\sim 10^6$ K, in the post-shock zone  until the abrupt drop at  $N_{\rm ion} \approx 10^{15}$ cm$^{-2}$, which is triggered by a thermal instability. Because the ionization time scale is similar to the flow time scale for many ions, the ionization increases from the pre-shock ionization behind the shock until the thermal instability. The temperature here drops by a factor of $\gtrsim 10^2$ until it is stabilized at $1000-1500$ K, where cooling is balanced by photoionization heating from the hot region.  At the same time the degree of ionization decreases and this region is dominated by neutral elements, although there is also a substantial fraction of singly and  doubly ionized ions. Because the whole post-shock region is close to pressure equilibrium, the density varies roughly the opposite way to the temperature (Fig. \ref{fig:pwn_shock}A),  with an ion density $\gtrsim 10^6 \ccm$. A magnetic pressure reduces this compression. The thickness of the shocked region is  $\sim 10^{11}$ cm, which would produce a narrow velocity line if the shock has a limited extent in solid angle. 

This model reproduces the observed relative luminosities of the [Ar~{\sc vi}],  [Ar~{\sc iii}], and  [S~{\sc iv}] lines, within the uncertainties. However,  the predicted [S~{\sc iii}] 18.70~{\textmu}m line is too faint by more than an order of magnitude, caused by the rapid recombination to lower ionization stages after the thermal instability. This may be a problem for this model.
In addition, the model predicts fairly strong [Si~{\sc i}] 1.607, 1.645~{\textmu}m lines as a result of the  high temperature in the Si~{\sc i} region (Fig. \ref{fig:pwn_shock}). 

This shock model predicts strong lines in the UV, including O~{\sc v}] 1218 \AA, Si~{\sc iv} 1394, 1403 \AA, O~{\sc iv}] 1401, 1408 \AA, and S~{\sc iv}] 1405-1424 \AA, and in the optical [O~{\sc iii}]   4959, 5007 \AA, [Ar~{\sc v}] 6435, 7006  \AA,  [Ca~{\sc ii}] 7291, 7323 \AA,  [Ar~{\sc iii}] 7136, 7751 \AA.  Dust scattering and absorption could decrease the fluxes. Unfortunately, the latest spectral HST observations from 2017--2018 \cite{Kangas2022},  do not have sufficient S/N or spectral resolution for a meaningful comparison. 

In summary, the PWN shock models have both pros and cons. They can  explain the velocity offset and narrow FWHM of the Ar lines, in particular if emission is from an anisotropic PWN bubble [as in SNR G54.1+0.3 \cite{Temim2010}]. 
The PWN scenario does not require dust absorption to explain the lack of lines above $\sim 8 \ \mu$m, although dust is known to be present in SN 1987A \cite{Matsuura2011,Cigan2019}.
However the observations indicate a lower density and higher shock velocities compared to the predictions of PWN model, although a clumpy medium could explain that difference. The most serious discrepance is the faint MIR [S~{\sc iii}] 18.70~{\textmu}m line in the model. In addition, the predicted [Si~{\sc i}] 1.607, 1.645~{\textmu}m lines in the NIR range are not seen, although this is similar to the CNS case (section \ref{sec:model_results}), and the same caveats, connected to the observations, ionization from the ER and ejecta and dust depletion, applies here.  Given the uncertainties in the atomic data, model parameters, geometry etc. we cannot rule out the PWN shock scenario completely. 

\subsection{Constraints on the NS kick velocity}
\label{sec:ns_kicks}
The position and velocity of the high ionization emission can in some scenarios provide a limit on the NS kick, which has been predicted by core collapse simulations \cite{Janka2017b}.  

The  centroid of the [Ar~{\sc vi}] position is $38 \pm 22$ mas east and $31 \pm 22$ mas south of the geometric center of the ER (Sect. \ref{sec:spatial_ar}). If we assume the ER center was the explosion site \cite{Alp2018}, over 35 years that translates to $249 \pm 146 \kms$ east and $208 \pm 146 \kms$ south of the ER center, assuming this to be the explosion site, or a combined velocity of  $324 \pm 206 \kms$  on the plane of the sky. Adding the radial velocity of $259.6 \pm 0.4 \kms$ (Table\ref{table:1}) in quadrature we get a total velocity of 
$416 \pm 206 \kms$ relative to the explosion position for the [Ar~{\sc ii, vi}]  emitting gas.

We compare this position and radial velocity to simulations of SN 1987A \cite{Ono2020,Orlando2020}, which predicted a kick velocity of $\sim -300 \kms$ in the line of sight, and to the north of the center. 
While the radial velocity of the  [Ar~{\sc ii}] line is similar to the predictions, the position of the [Ar~{\sc vi}] emission differs substantially.

The kick velocity of the NS depends on the  relative positions of the NS and the emitting gas in the different scenarios. In the PWN case the emission would be exterior to or in the shocked shell, separating the bubble of relativistic particles and the ejecta (Sect. \ref{sec:origin}). The observed radial velocity of $\sim -260 \kms$, as well as sky position, are therefore indicative of the line emitting site, rather than that of the NS. 

For a low magnetic field and/or a long pulsar period the non-thermal energy input from the NS would be low. In particular, the class of NSs known as central compact objects (CCOs) in young SNRs, which represent a large fraction of young pulsars, have magnetic fields, $10^{10}-10^{11}$ G,  and spin periods $0.3 \pm 0.15$ s \cite{DeLuca2017}. In that case, the dynamic effects on the surrounding gas of the NS would be small, 
so the emitting gas could be closer to the NS, so the transverse   and radial velocity above might approximate the kick velocity. The FWHM of the [Ar~{\sc ii}] line, $\sim 122 \kms$, might then indicate the size of the region ionized by the CNS, $\sim 1.3 \times 10^{16}$ cm, which is within an order of magnitude of the size of the ionized region in the CNS model. This scenario could explain why only a blue shift is observed, rather than a more boxy line profile, as would be expected for a spherically symmetric shell without absorption.

\end{document}